\documentclass[10pt,twocolumn,twoside]{IEEEtran}
\usepackage{amsmath,amssymb,amsfonts}
\usepackage{cite}
\usepackage{graphicx}
\usepackage{amsmath}
\usepackage{array}
\usepackage{algorithmicx}
\usepackage{algpseudocode}
\usepackage{setspace}
\usepackage{blindtext}
\usepackage{url}
\usepackage{epstopdf}
\usepackage{verbatim}
\usepackage{color}
\usepackage{amsmath}
\usepackage{bm}
\usepackage{multirow}
\usepackage[table,xcdraw]{xcolor}
\usepackage{booktabs}
\usepackage{makecell}
\usepackage{ntheorem}
\usepackage{textcomp}
\usepackage{amsmath}
\usepackage{mathrsfs}
\usepackage{graphicx}
\usepackage{subcaption}
\usepackage[utf8]{inputenc}
\usepackage[font=small,labelfont=bf]{caption}
\usepackage{tikz}
\usetikzlibrary{arrows.meta, positioning, shapes, calc, fit}
\usetikzlibrary{decorations.pathreplacing}
\makeatletter
\newcommand{\gettikzxy}[3]{%
  \tikz@scan@one@point\pgfutil@firstofone#1\relax
  \edef#2{\the\pgf@x}%
  \edef#3{\the\pgf@y}%
}
\usetikzlibrary{spy,backgrounds}

\newtheorem{Prob}{Problem}

\newcommand{\RNum}[1]{\uppercase\expandafter{\romannumeral #1\relax}}

\theoremseparator{:}

\bstctlcite{IEEEexample:BSTcontrol}
\begin{document}

\title{A Retrieval-Assisted Framework for Wireless Localization}

\author{
Haoyu Huang, Guangjin Pan, \IEEEmembership{Member, IEEE}, Kaixuan Huang,                                Shunqing Zhang, \IEEEmembership{Senior Member, IEEE}, \\
Yuhao Zhang, Musa Furkan Keskin, \IEEEmembership{Member, IEEE}, Zheng Xing, \IEEEmembership{Member, IEEE}, Henk Wymeersch, \IEEEmembership{Fellow, IEEE}
\thanks{This work was supported by the Science and Technology Commission Foundation of Shanghai under Grant 24DP1500703, and the National Natural Science Foundation of China (NSFC) under Grant 62571307. The work of Guangjin Pan, Yuhao Zhang, Musa Furkan Keskin, and Henk Wymeersch was supported in part by the SNS JU project 6G-DISAC
under the EU’s Horizon Europe research and innovation Program under Grant
Agreement No 101139130, the Swedish Foundation for Strategic Research
(SSF) (grant FUS21-0004, SAICOM),  the Swedish Research Council (VR) through the project 6G-PERCEF under Grant 2024-04390, and the European Union through the project ISLANDS under Grant 101120544.
(Corresponding authors: Guangjin Pan, Shunqing Zhang)}
\thanks{Haoyu Huang, Kaixuan Huang, and Shunqing Zhang are with the School of Communication and Information Engineering, Shanghai University, 200444 Shanghai, China (e-mail: {hhy\_123; xuan1999; shunqing}@shu.edu.cn).}
\thanks{Guangjin Pan, Yuhao Zhang, Musa Furkan Keskin, and Henk Wymeersch are with the Department of Electrical Engineering, Chalmers University of Technology, 41296 Gothenburg, Sweden (e-mail: {guangjin.pan; yuhaozh; furkan; henkw}@chalmers.se)}
\thanks{Zheng Xing is College of Computer Science and Software Engineering, Shenzhen University, Shenzhen, 518060, Guangdong, China (e-mail: zhengx@szu.edu.cn)}
}

\IEEEpeerreviewmaketitle

\maketitle

\begin{abstract}
Accurate and robust wireless localization is a key enabler for a wide range of mobile computing applications. 
Fingerprint-based localization using channel state information (CSI) has attracted significant attention due to its high accuracy and compatibility with existing communication infrastructures. 
However, traditional similarity-based fingerprinting methods suffer from high computational complexity and limited scalability in high-dimensional CSI spaces, while purely learning-based approaches fail to explicitly exploit correlations among reference fingerprints during inference.
To address these challenges, this paper proposes a unified retrieval-assisted fingerprinting localization framework that tightly integrates similarity-based and learning-based paradigms. 
Specifically, channel charting is employed to project high-dimensional CSI into a low-dimensional latent space, enabling efficient and scalable retrieval of locally correlated reference points (RPs). 
Building upon the retrieved RPs, a graph attention network (GAT) is designed to explicitly model inter-sample correlations between the query CSI and its associated references, allowing adaptive and geometry-aware feature aggregation for accurate position estimation.
Extensive experiments conducted on both real-world indoor and ray-tracing simulated outdoor scenarios demonstrate that the proposed method consistently outperforms state-of-the-art similarity-based and learning-based localization approaches.
\end{abstract}


\section{Introduction} \label{sect:intro}

Accurate and robust wireless localization is a fundamental enabler for a wide range of mobile computing applications, including indoor navigation, industrial automation, intelligent transportation systems, and extended reality (XR)~\cite{pan2025ai, Zafari2019indoor,Henk2025 }. Beyond individual applications, localization serves as a core contextual awareness capability that enables intelligent interaction between mobile devices, users, and their surrounding environments. 
With the evolution toward 6G networks, localization is expected to be deeply integrated into communication systems and provided as a native network capability rather than an auxiliary service, supporting emerging location-aware services and intelligent network functions~\cite{yang2024positioning}.

Geometry-based wireless localization methods estimate user positions by leveraging measurable channel parameters, e.g., time-of-arrival (TOA), time-difference-of-arrival (TDOA) and angle-of-arrival (AOA)~\cite{8515231,Yang2024}. Although these approaches are theoretically well established, they typically rely on strict assumptions, e.g., tight synchronization, accurate anchor positions, and reliable line-of-sight (LOS) conditions~\cite{Italiano5GTutorial}. In practical environments characterized by non-line-of-sight (NLOS) propagation and rich multipath effects, these assumptions are often violated, resulting in degraded localization accuracy and limited robustness~\cite{zhang2025leveraging}. To address these challenges, fingerprint-based localization methods associate signal measurements, e.g., channel state information (CSI), with a pre-collected fingerprint–location database, and have emerged as a promising paradigm due to their high accuracy and compatibility with existing communication infrastructures~\cite{chen2022tutorial}.

Traditional fingerprinting approaches typically rely on similarity-based matching schemes, such as K-nearest neighbors (KNN) and its variants, where the location of a query user is inferred from the most similar fingerprints in a pre-collected database~\cite{Sadowski_2020_KNN,KNN2022Ali,Hu_2024_KNN,luckner2023selection}. For example, in \cite{Hu_2024_KNN}, the authors propose a multi-band multi-cell fingerprint combined with an enhanced KNN algorithm to improve outdoor localization accuracy. The work in \cite{luckner2023selection} investigates how access point (AP) selection strategies can optimize the computational efficiency of KNN in fingerprint-based localization by reducing feature dimensionality while maintaining accuracy.
However, in high-dimensional CSI feature spaces, similarity-based matching requires exhaustive pairwise distance computations between the query sample and all reference fingerprints, resulting in high computational complexity and poor scalability. This issue becomes particularly critical in large-scale deployments or real-time mobile computing scenarios, where the size of the fingerprint database continuously grows and low-latency localization is required \cite{yang2024positioning}.

To overcome these limitations, learning-based localization methods have been extensively studied, where deep neural networks are trained to directly map CSI measurements to physical coordinates. By exploiting the strong representational capability of deep models, these approaches can capture complex nonlinear relationships between channel characteristics and spatial locations, enabling localization during inference based solely on real-time collected fingerprints. 
A variety of neural network architectures have been explored in this context, including multilayer perceptrons (MLPs) \cite{Lin2025_MLP}, convolutional neural networks (CNNs) \cite{GAO_2024CNN}, recurrent neural networks \cite{zhang2023csi}, transformer \cite{pan2025large}, and graph neural networks (GNNs) \cite{Zhang2023_GNN1,Ye_2025_GNN2,GCNLoc}. These approaches focus on improving channel representation learning for localization tasks by using increasingly expressive neural architectures or training methods. For example, the authors in \cite{pan2025large} build upon a transformer-based localization framework and employ self-supervised learning to improve channel representation learning, leading to improved positioning accuracy.
The work in \cite{Zhang2023_GNN1} employs GNNs to model WiFi fingerprint data as graphs, explicitly capturing the relationships between APs and UEs to enhance channel representation. \cite{Ye_2025_GNN2} proposes a hierarchical GNN framework that constructs intra-AP and inter-AP graphs and incorporates unsupervised domain adaptation to improve robustness against obstacles and environmental changes.
Despite their effectiveness, purely learning-based localization solutions suffer from several inherent drawbacks.
In fingerprint-based localization, purely learning-based methods implicitly capture the spatial distribution of wireless fingerprints by training on labeled datasets to establish mappings between channel characteristics and physical coordinates \cite{pan2025ai}. However, these approaches inherently rely on substantial volumes of annotated CSI-position pairs to achieve robust performance. In scenarios with scarce labeled data, which is common in practice due to costly data collection and calibration, such methods exhibit significant performance degradation, including reduced positioning accuracy and compromised robustness against environmental variations \cite{Zhu2023} .

Another localization method is based on channel charting (CC) \cite{stahlke2023indoor,stephan2024angle,ferrand2021triplet,taner2025channel,mateos2025positioning,zhang2025unilocpro}. By exploiting the correlations among CSI samples, channel charting employs self-supervised learning to compress high-dimensional CSI into low-dimensional representations while preserving the intrinsic geometric structure of the radio environment \cite{pan2025ai}. As a result, CC inherits the representation learning capability of learning-based localization methods while simultaneously capturing similarity-preserving channel characteristics. These learned representations have been shown to be effective in improving localization accuracy by facilitating geometry-aware similarity operations. While some works focus on enhancing geometric properties within the channel chart itself~\cite{stahlke2023indoor,stephan2024angle,ferrand2021triplet}, others aim to directly embed the chart into real-world coordinates~\cite{taner2025channel,mateos2025positioning,zhang2025unilocpro}.
For example, the authors in \cite{stephan2024angle} enhance the channel geometric similarity by leveraging the channels in the angle-delay domain, improving localization accuracy under NLOS conditions.\cite{taner2025channel} proposes a weakly-supervised framework that embeds channel charts into real-world coordinates by incorporating geometric constraints, such as known AP positions and estimated LOS regions. The work in \cite{zhang2025unilocpro} introduces novel dissimilarity metrics to accurately capture the geometric relationships of channel paths, achieving accurate localization under mixed NLoS conditions.  
However, despite existing CC methods typically adopting similarity-based self-supervised learning during training, localization based on CC heavily relies on the dissimilarity metrics employed, and it is often challenging to identify dissimilarity metrics that exhibit a direct correlation with physical distance. Moreover, their inference stage still operates on individual CSI samples, failing to explicitly exploit the rich information in the pre-collected dataset, thereby limiting achievable performance gains from data-driven fingerprinting systems.

This paper proposes a unified retrieval-assisted fingerprinting localization framework that tightly integrates similarity-based and learning-based localization paradigms. Specifically, we employ CC to transform high-dimensional CSI into a low-dimensional latent space, enabling efficient retrieval of locally correlated reference points (RPs), i.e., pre-collected CSI samples with known positions that exhibit high similarity to the query sample. Building upon the retrieved RPs, a graph is dynamically constructed for each query, where nodes represent the query CSI and its retrieved RPs. A graph attention network (GAT) is then applied to explicitly model the pairwise correlations within this graph, enabling adaptive feature aggregation from the most relevant references for accurate position estimation. The main
contributions of our work are summarized as follows:
\begin{itemize}
    \item \textbf{A unified retrieval-assisted localization framework:} We propose a novel two-stage localization architecture that tightly integrates RP retrieval with learning-based localization inference. This framework establishes a new paradigm by bridging similarity-based fingerprinting and deep learning–based positioning, enabling explicit exploitation of cross-sample correlations during inference.

    \item \textbf{CC–GAT-enabled localization algorithm:}  We develop a CC–assisted retrieval mechanism that learns geometry-preserving low-dimensional representations of high-dimensional CSI in a self-supervised manner, significantly reducing retrieval complexity while maintaining high-quality RP selection. Building upon the retrieved references, we further design a GAT that explicitly models the relational dependencies between the query CSI and its associated RPs, enabling adaptive and robust feature aggregation for accurate localization.

    \item \textbf{Comprehensive evaluation on real-world and simulated datasets:} 
    Extensive experiments on both real-world indoor measurement and ray-tracing-based outdoor simulation datasets demonstrate that the proposed framework consistently outperforms state-of-the-art localization methods, particularly in few-shot and data-scarce scenarios. 
    With 1000 fully labeled samples, our proposed framework achieves mean localization errors of $0.8 \, m$ on the indoor dataset and $3.56 \, m$ on the outdoor dataset, representing improvements of 50.6\% and 53.8\% compared to baseline methods, respectively. Furthermore, within the proposed framework, the GAT-based network outperforms other architectures, achieving an improvement of 13.9\% in mean localization errors.
\end{itemize}

The remainder of this paper is organized as follows. Sec. \ref{sect:sys} introduces the system model and presents the proposed retrieval-assisted localization framework. Sec. \ref{sect:solution} details the CC–GAT–based solution. Sec. \ref{sect:Experiment} reports experimental results and performance evaluations. Finally, Sec. \ref{sect:conclusion} concludes the paper.

\section{System Model} \label{sect:sys}

We consider an uplink wireless localization system comprising $N_{\text{BS}}$ base stations (BSs), where each BS is equipped with a uniform linear array (ULA) of $N_r$ receive antennas and employs orthogonal frequency-division multiplexing (OFDM) with $N_{\text{SC}}$ subcarriers.
Each user equipment (UE) is equipped with a single transmit antenna and moves within a two-dimensional spatial domain\footnote{The proposed framework is generalizable to three-dimensional localization, all experiments in this paper focus on two-dimensional scenarios for consistency with the evaluation datasets.}, periodically transmitting pilot symbols that are received simultaneously by multiple BSs for localization purposes.

\subsection{Channel Model}
\label{subsect:channel}

At a given user position, the uplink channel between the UE and the $b$-th BS is modeled as the superposition of $N_{\text{path},b}$ multipath components (MPCs). 
For the $m$-th receive antenna and the $l$-th subcarrier, 
the complex channel coefficient observed at BS $b$ is expressed as
\begin{equation}
h_{b,m,l} =
\sum_{p=1}^{N_{\text{path},b}}
\beta_{p,b}
e^{-j 2\pi f_l \tau_{p,b}}
[\bm{a}_b(\varphi_{p,b})]_m
+ n_{b,m,l},
\label{eq:channel_multiBS}
\end{equation}
where $\beta_{p,b}$ denotes the complex gain of the $p$-th path observed at BS $b$, 
$\tau_{p,b}$ is the corresponding propagation delay 
and $\varphi_{p,b}$ represents the AoA at BS $b$. 
The subcarrier frequency is given by $f_l = f_c + (l-1)\Delta_f$, 
and $n_{b,m,l}$ denotes additive white Gaussian noise (AWGN). 
The vector $\bm{a}_b(\cdot)$ represents the array response of BS $b$, 
capturing the phase offsets among antenna elements for a given AoA $\varphi$. 
For a ULA with inter-element spacing $d$, 
it is defined as
\begin{equation}
\bm{a}_b(\varphi) =
\left[
1,~
e^{-j\frac{2\pi d}{\lambda}\sin(\varphi)},~
\dots,~
e^{-j\frac{2\pi d}{\lambda}(N_r - 1)\sin(\varphi)}
\right]^\top,
\label{eq:array_response_multiBS}
\end{equation}
where $\lambda$ denotes the carrier wavelength.

By collecting all antenna responses across the $N_{\text{SC}}$ subcarriers, 
the overall frequency-domain channel matrix observed at the $b$-th BS can be written as \cite{Yuan2025}
\begin{equation}
\bm{H}_b =
[\bm{h}_{b,1}, \bm{h}_{b,2}, \dots, \bm{h}_{b,N_{\text{SC}}}]
\in \mathbb{C}^{N_r \times N_{\text{SC}}},
\label{eq:CSI_matrix_multiBS}
\end{equation}
where $\bm{h}_{b,l} = [h_{b,1,l}, \dots, h_{b,N_r,l}]^\top$ 
represents the channel frequency response (CFR) vector of the $l$-th subcarrier at BS $b$. The frequency-domain channel matrix $\bm{H}_b$ can be transformed into its delay-domain representation 
by applying the inverse discrete Fourier transform (IDFT) along the subcarrier dimension \cite{stephan2024angle}. 
The resulting delay-domain channel impulse response (CIR), denoted by $\tilde{\bm{H}}_b$, 
reveals the multipath delay profile observed at BS $b$.  By aggregating the CIR $\tilde{\bm{H}}_b$ from all BSs, the multi-BS frequency-domain channel representation can be organized as a tensor $\mathcal{H} \in \mathbb{C}^{N_{\text{BS}} \times N_r \times N_{\text{SC}}}$.

\subsection{Dataset Model}
\label{subsect:dataset}

AI–driven localization relies on datasets collected from realistic or simulated environments to enable model training~\cite{pan2025ai}. 
In this work, we construct a labeled dataset consisting of multi-BS CSI tensors and their corresponding ground-truth user positions:
\begin{equation}
\mathcal{D}_{\text{lab}} 
= \big\{ \, (\mathcal{H}_i, {\mathcal{L}}_i) 
~\big|~ 
i = 1, 2, \dots, N_{\text{lab}} \, \big\},
\label{eq:dataset}
\end{equation}
where $\mathcal{H}_i \in \mathbb{C}^{N_{\text{BS}} \times N_r \times N_{\text{SC}}}$ 
denotes the aggregated channel tensor observed across all BSs for the $i$-th sample, 
and ${\mathcal{L}}_i \in \mathbb{R}^{2}$ 
represents the corresponding user position in the two-dimensional spatial domain.
The total number of CSI-position pairs in the dataset is denoted by $N_{\text{lab}}$.

\subsection{Preliminaries of Data-driven Localization}
\label{subsect:prelim}

In data-driven wireless localization, CSI serves as a high-dimensional feature that implicitly captures the spatial characteristics of the radio environment. 
Each CSI sample $\mathcal{H}_i$ recorded at position $\mathcal{L}_i$ carries a unique propagation signature that can be used to infer the user’s location. 
The fundamental objective of localization is to establish a mapping between the observed CSI and the corresponding spatial coordinates. Typically, fingerprint-based localization methods can be categorized into two main classes: 

\subsubsection{Similarity-based localization} 
In similarity-based localization, the dataset $\mathcal{D}_{\text{lab}}$ serves as a reference database that stores CSI fingerprints associated with known positions. 
Given an unknown CSI measurement $\mathcal{H}_j$, the user’s position is estimated by retrieving the most similar RPs within $\mathcal{D}_{\text{lab}}$. 
A representative example is the weighted $K$–nearest neighbors (WKNN) algorithm \cite{zhou2021integrated}, where the estimated location is obtained as the weighted average of the coordinates corresponding to the $K$ most similar fingerprints \cite{hu2018experimental}:
\begin{equation}
\widehat{\mathcal{L}}_j = 
\frac{ \sum_{(\mathcal{H}_i,\mathcal{L}_i)\in \mathcal{D}_{\text{ref}}(\mathcal{H}_j)} w_i \, \mathcal{L}_i}
     { \sum_{(\mathcal{H}_i,\mathcal{L}_i)\in \mathcal{D}_{\text{ref}}(\mathcal{H}_j)} w_i},
\label{eq:wknn}
\end{equation}
where $\mathcal{D}_{\text{ref}}(\mathcal{H}_j)$ denotes the subset of $\mathcal{D}_{\text{lab}}$ containing the $K$ RPs most similar to the query CSI $\mathcal{H}_j$, i.e.,
\begin{equation}
\mathcal{D}_{\text{ref}}(\mathcal{H}_j) = \mathcal{G}(\mathcal{D}_{\text{lab}}, \mathcal{H}_j),
\label{eq:G_def}
\end{equation}
and $\mathcal{G}(\cdot)$ represents the retrieval function that returns the $K$ nearest fingerprints from $\mathcal{D_{\text{ref}}}$ according to a distance metric $d(\cdot,\cdot)$, which quantifies the dissimilarity between two CSI samples.
Here, $w_i$ denotes the similarity weight assigned to the $i$-th reference point, reflecting its relevance to the query sample.
A common definition for the similarity weight $w_i$ is given by \cite{hu2018experimental}:
\begin{equation}
w_i = \frac{1}{d(\mathcal{H}_j, \mathcal{H}_i)},
\label{eq:weights}
\end{equation}
though alternative weighting schemes exist in the literature. The distance metric $d(\cdot,\cdot)$ can be the Euclidean distance, correlation distance, or any other appropriate similarity measure.

\subsubsection{Learning-based localization} 
Learning-based methods aim to train a parametric model that directly maps CSI features to spatial coordinates. 
Specifically, a model $\mathcal{F}(\cdot)$ with learnable parameters $\bm{\theta}$ is optimized to minimize the mean-squared localization error:
\begin{equation}
\bm{\theta}^\ast = 
\arg \min_{\bm{\theta}}
\frac{1}{N_{\text{lab}}}
\sum_{i=1}^{N_{\text{lab}}}
\left\|
\mathcal{L}_i - \mathcal{F}(\mathcal{H}_i;\bm{\theta})
\right\|_2^2.
\label{eq:learning_loss}
\end{equation}
Deep neural networks (DNNs)  have been widely adopted as implementations of $\mathcal{F}(\cdot)$\cite{yang2024positioning}, enabling nonlinear CSI-to-location mappings and improved generalization across environments. 
Once trained, the model performs inference solely based on its internal parameters, without explicitly exploiting the correlations between the query CSI sample and the RPs stored in the database.

\subsection{Proposed Localization Framework}
\label{sect:hybrid}

\begin{figure}[t]
    \centering
\includegraphics[scale=0.35]{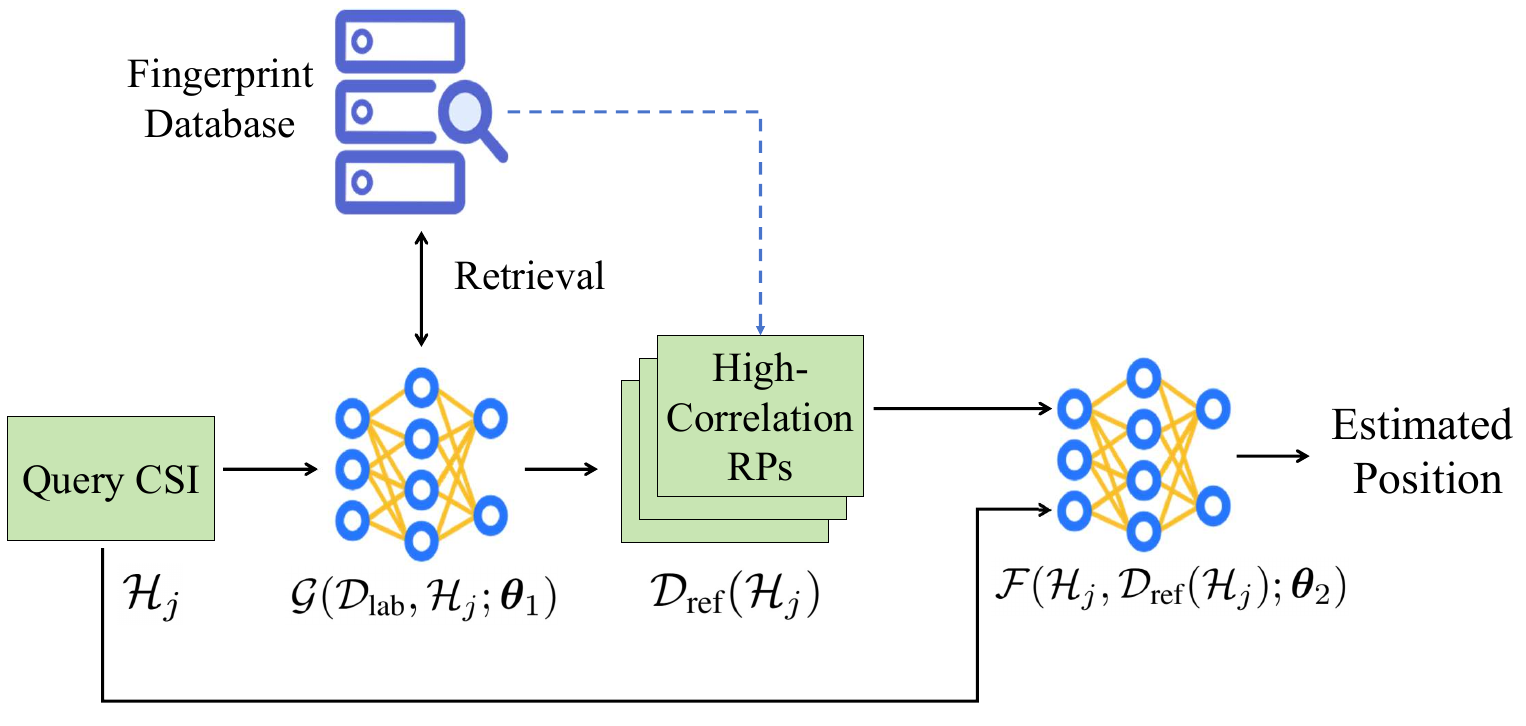}
    \caption{
        The proposed retrieval-assisted fingerprinting localization framework, 
        which combines the advantages of similarity-based and learning-based localization.
    }
    \label{fig:framework}
\end{figure}

To address the limitation of data-driven localization, we propose a hybrid paradigm that integrates learning-based and similarity-based approaches, where retrieved RPs provide relevant information to assist inference. Such integration combines the generalization capability of deep learning with the robustness of similarity search. To combine the advantages of similarity-based and learning-based localization, 
we propose a retrieval-assisted fingerprinting localization framework, as illustrated in Fig.~\ref{fig:framework}. 
The framework operates in two main stages:
\begin{itemize}
    \item \textbf{RP retrieval:} 
    Given a CSI tensor $\mathcal{H}_j$ collected at an unknown position, 
    the reference subset $\mathcal{D}_{\text{ref}} (\mathcal{H}_j)$ is retrieved from the database $\mathcal{D}$ 
    using a retrieval function $\mathcal{G}(\cdot\,;\bm{\theta}_1)$ which is implemented as a neural network, i.e.,
    \begin{equation}
        \mathcal{D}_{\text{ref}}(\mathcal{H}_j) = 
        \mathcal{G}(\mathcal{D}_{\text{lab}}, \mathcal{H}_j; \bm{\theta}_1),
        \label{eq:ref_set}
    \end{equation}
    where $\bm{\theta}_1$ denotes the parameters of the retrieval module that identifies the most relevant RPs for the query sample.

    \item \textbf{Localization inference:} 
    The retrieved subset $\mathcal{D}_{\text{ref}}(\mathcal{H}_j)$ is then combined with the query CSI tensor $\mathcal{H}_j$ 
    and fed into a localization network $\mathcal{F}(\cdot\,;\bm{\theta}_2)$ for position estimation:
    \begin{equation}
        \widehat{\mathcal{L}}_j = 
        \mathcal{F}(\mathcal{H}_j, \mathcal{D}_{\text{ref}}(\mathcal{H}_j); \bm{\theta}_2),
        \label{eq:loc_func}
    \end{equation}
    where $\bm{\theta}_2$ represents the learnable parameters of the localization model.
\end{itemize}
By incorporating both the query CSI and its most relevant RPs, the proposed retrieval-assisted framework enables the network to leverage cross-sample correlations that capture the underlying geometric and multipath characteristics of the radio environment, thereby improving localization accuracy, robustness, and generalization capability.

Based on the proposed framework, the localization task can be formulated as an optimization problem that minimizes the expected positioning error between the estimated and ground-truth user locations:
\begin{Prob}[\em Localization Error Minimization]
The overall localization error minimization problem for the proposed retrieval-assisted fingerprinting localization framework is formulated as
\begin{eqnarray}
\min_{\bm{\theta}_1,\bm{\theta}_2} & & 
\frac{1}{|\mathcal{D}_{\text{lab}}|}\sum_{\{\mathcal{H}_j, \mathcal{L}_j\} \in \mathcal{D}_{\text{lab}}} 
\|\mathcal{L}_j - \widehat{\mathcal{L}}_j\|_2, 
\label{eqn:overall_min}\\
\mathrm{s.t.} & & 
\mathcal{D}_{\text{ref}}(\mathcal{H}_j) = 
\mathcal{G}(\mathcal{D}_{\text{lab}}, \mathcal{H}_j; \bm{\theta}_1),
\label{eqn:ref_func}\\
& & 
\widehat{\mathcal{L}}_j = 
\mathcal{F}(\mathcal{H}_j, \mathcal{D}_{\text{ref}}(\mathcal{H}_j); \bm{\theta}_2), 
\label{eqn:loc_func}
\end{eqnarray}
where $\|\cdot\|_2$ represents the $\ell_2$ norm of the localization error, 
$\mathcal{L}_j$ and $\widehat{\mathcal{L}}_j$ denote the ground-truth and estimated positions of the $j$-th sample, respectively. 
The retrieval module $\mathcal{G}(\cdot\,;\bm{\theta}_1)$ selects the most relevant RPs from the labeled dataset $\mathcal{D}_{\text{lab}}$, 
while the localization network $\mathcal{F}(\cdot\,;\bm{\theta}_2)$ leverages both the query CSI and the retrieved references for position estimation. 
$\mathcal{D}_{\text{lab}}$ serves both as the retrieval repository and the source of training queries.

\end{Prob}

Implementing the proposed framework involves two key challenges.
The first is to design an effective retrieval function $\mathcal{G}(\cdot\,;\bm{\theta}_1)$ that selects a reference subset $\mathcal{D}_{\text{ref}}(\mathcal{H}_j)$ highly correlated with the query CSI $\mathcal{H}_j$, ensuring that the retrieved fingerprints provide informative spatial and multipath cues for localization. 
The second challenge lies in constructing a localization network $\mathcal{F}(\cdot\,;\bm{\theta}_2)$ that can adaptively aggregate information from the retrieved reference set $ \mathcal{D}_{\text{ref}}(\mathcal{H}_j)$ and the query sample $\mathcal{H}_j$  to form correlation-aware representations for accurate and robust position estimation.

\section{CC-GAT-based solution}
\label{sect:solution}

\begin{figure*}
\centering
\includegraphics[scale=0.7]{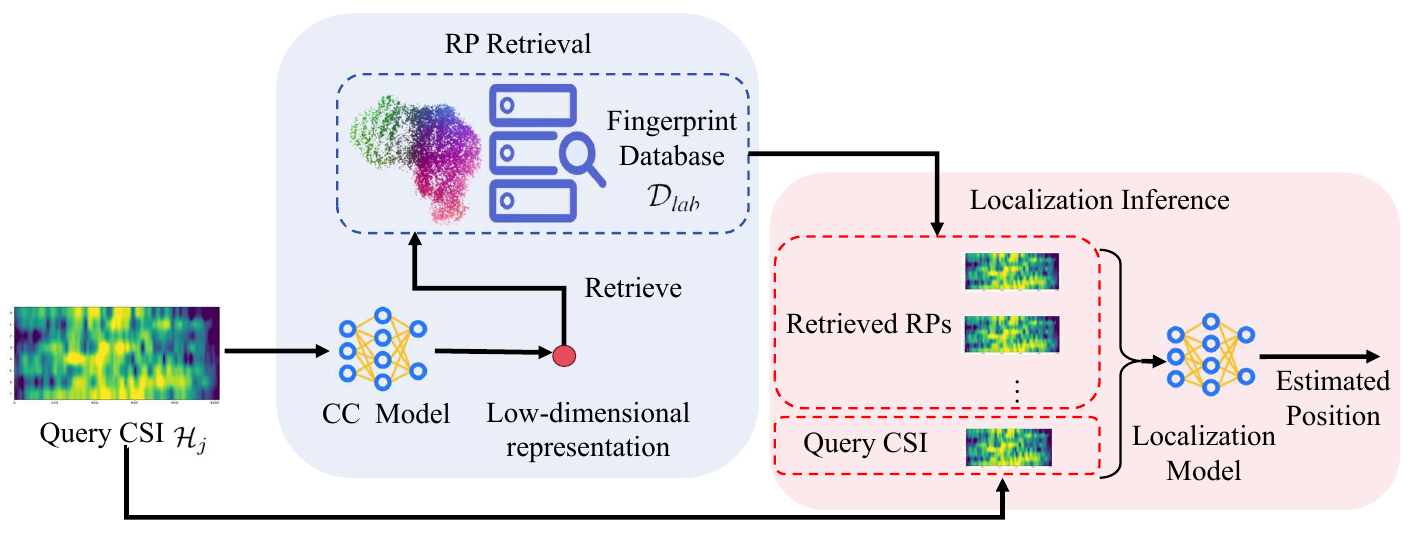}
\caption{Architecture of the proposed CC-GAT-based localization framework, where channel charting is used for efficient RP retrieval and a GAT performs correlation-aware localization.}
\label{fig:fig2}
\end{figure*}

To efficiently solve the optimization problem defined in Section~\ref{sect:hybrid}, 
we propose a CC-assisted GAT solution. The overall architecture of the proposed solution is illustrated in Fig.~\ref{fig:fig2}:
\begin{itemize}
    \item For the retrieval function $\mathcal{G}(\cdot\,;\bm{\theta}_1)$, a straightforward approach is to directly compute the pairwise similarity between the query CSI $\mathcal{H}_j$ and all fingerprints $\{\mathcal{H}_i\}$ in the database $\mathcal{D}_{\text{lab}}$, using metrics such as the Euclidean distance. However, this exhaustive search becomes computationally prohibitive in high-dimensional CSI spaces. To address this issue, we employ channel charting \cite{stephan2024angle} to obtain a low-dimensional manifold representation of the CSI space. By applying dimensionality reduction techniques such as autoencoder \cite{stephan2024angle}, Siamese network \cite{taner2025channel} and triplet method \cite{ferrand2021triplet}, channel charting preserves the intrinsic geometric relationships among CSI samples in a compact latent space.
    Within this low-dimensional chart, similar RPs can be efficiently retrieved through simple distance computations, thereby significantly reducing computational complexity. Although the RPs retrieved via channel charting may not be globally optimal for the positioning task, they effectively retain local geometric similarity, thus providing a reliable and low-complexity foundation for subsequent localization.

    \item To further exploit the spatial correlations between the query CSI and the retrieved RP set, we employ a GAT for the localization module $\mathcal{F}(\cdot\,;\bm{\theta}_2)$\cite{velickovic2017graph}. GAT enables adaptive feature aggregation by learning attention weights that reflect the relevance of each reference fingerprint to the query sample. By explicitly modeling inter-sample correlations, this mechanism enhances geometry-aware representation learning, leading to more accurate and robust position estimation.

\end{itemize}
In the following, we provide detailed descriptions of the two key components of the proposed framework.

\subsection{Channel Charting}
\label{subsect:CC}

Several representative CC approaches have been proposed in the literature, 
including autoencoder-based methods, Siamese networks, and triplet-based learning frameworks. 
Unlike conventional supervised localization, 
CC typically adopts a self-supervised or weakly supervised paradigm 
and therefore does not require explicit position labels during training. 
Given an unlabeled CSI dataset 
$\mathcal{D}_{\text{ul}} = \{\mathcal{H}_i \mid i = 1, 2, \dots, N_{\text{ul}}\}$, 
where $N_{\text{ul}}$ denotes the total number of unlabeled CSI samples, 
we introduce three representative CC methods in the following.

\subsubsection{Autoencoder}
Autoencoder-based channel charting \cite{stephan2024angle} aims to learn a nonlinear mapping 
from high-dimensional CSI data to a compact latent representation through unsupervised learning. 
An autoencoder consists of an encoder $\mathcal{G}_{\text{enc}}^{\text{ae}}(\cdot; \bm{\theta}_e)$ 
and a decoder $\mathcal{G}_{\text{dec}}^{\text{ae}}(\cdot; \bm{\theta}_d)$, where $\bm{\theta}_e$ and $\bm{\theta}_d$ denote the trainable parameters of the encoder and decoder, respectively.
These components jointly trained to minimize the reconstruction error between the input and its reconstructed output:
\begin{equation}
\mathcal{J}_{\text{AE}} = 
\frac{1}{N_{\text{ul}}}\sum_{\mathcal{H}_i \in \mathcal{D}_{\text{ul}}}
\left\|
\mathcal{H}_i -
\mathcal{G}_{\text{dec}}^{\text{ae}}
\big(
\mathcal{G}_{\text{enc}}^{\text{ae}}(\mathcal{H}_i; \bm{\theta}_e);
\bm{\theta}_d
\big)
\right\|_2^2.
\label{eq:ae_loss}
\end{equation}
Through this reconstruction objective, the encoder learns to extract 
a low-dimensional embedding $\bm{z}_i = \mathcal{G}_{\text{enc}}^{\text{ae}}(\mathcal{H}_i; \bm{\theta}_e)$, 
which represents the coordinate of the $i$-th sample in the channel chart. The dimensionality of $\bm{z}_i$ is typically small (e.g., 2 or 3) to preserve geometric structure while enabling efficient retrieval in the latent space.

\subsubsection{Siamese Network}

The Siamese network is a classical metric-learning framework that learns to preserve 
the relative similarity between pairs of CSI samples in the latent embedding space \cite{taner2025channel}. 
Instead of reconstructing the input data as in the autoencoder, 
the Siamese network is trained to learn and preserve a predefined CSI dissimilarity metric 
$d_{\text{CSI}}(\cdot,\cdot)$ between pairs of samples 
(e.g., Euclidean distance and correlation dissimilarity). 
The training objective of the encoder $\mathcal{G}_{\text{enc}}^{\text{siam}}(\cdot; \bm{\theta}_e)$ is formulated as:
\begin{equation}
\mathcal{J}_{\text{Siamese}}
= \sum_{\mathcal{H}_i,\mathcal{H}_j \in\mathcal{D}_{\text{ul}}} 
{\rho_{ij}}\,\Big(\,\|\mathbf{z}_i-\mathbf{z}_j\|_2 - d_{\text{CSI}}(\mathcal{H}_i,\mathcal{H}_j)\,\Big)^2,
\label{eq:cc_Siamese}
\end{equation}
where $\mathbf{z}_i=\mathcal{G}_{\text{enc}}^{\text{siam}}(\mathcal{H}_i; \bm{\theta}_e)$ 
denotes the latent embedding of CSI sample $\mathcal{H}_i$. 
The weighting coefficient ${\rho_{ij}}$ emphasizes local neighborhood preservation, 
which is typically set as ${\rho_{ij}}=1/d_{\text{CSI}}(\mathcal{H}_i,\mathcal{H}_j)$ \cite{taner2025channel}.

\subsubsection{Triplet-based Method}
The triplet-based method extends the Siamese framework by modeling relative relationships among triplets of CSI samples, 
consisting of an anchor sample, a positive sample collected at a nearby location, 
and a negative sample from a distant location \cite{ferrand2021triplet}. 
The key objective is to ensure that the anchor and positive samples are closer to each other 
than the anchor and negative samples in the latent embedding space. 
For a given anchor CSI sample $\mathcal{H}_a$, 
its positive and negative samples $\mathcal{H}_p$ and $\mathcal{H}_n$ are selected such that
\begin{equation}
d_{\text{CSI}}(\mathcal{H}_a, \mathcal{H}_p) 
< 
d_{\text{CSI}}(\mathcal{H}_a, \mathcal{H}_n),
\end{equation}
ensuring that $\mathcal{H}_p$ is more similar to $\mathcal{H}_a$ than $\mathcal{H}_n$.
Based on this selection, the encoder network 
$\mathcal{G}_{\text{enc}}^{\text{tri}}(\cdot; \bm{\theta}_e)$ 
is trained using the triplet loss:
\begin{equation}
\mathcal{J}_{\text{triplet}} 
= \sum_{(a,p,n) \in \mathcal{T}}
\big[
\|\mathbf{z}_a - \mathbf{z}_p\|_2^2 
- \|\mathbf{z}_a - \mathbf{z}_n\|_2^2 
+ Q
\big]_+,
\label{eq:triplet_loss}
\end{equation}
where 
$\mathbf{z}_a = \mathcal{G}_{\text{enc}}^{\text{tri}}(\mathcal{H}_a; \bm{\theta}_e)$, 
$\mathbf{z}_p = \mathcal{G}_{\text{enc}}^{\text{tri}}(\mathcal{H}_p; \bm{\theta}_e)$, 
and 
$\mathbf{z}_n = \mathcal{G}_{\text{enc}}^{\text{tri}}(\mathcal{H}_n; \bm{\theta}_e)$ 
represent the latent embeddings of the anchor, positive, and negative CSI samples, respectively. 
The margin parameter $Q$ enforces a minimum distance separation between dissimilar pairs, 
and $[\cdot]_+ = \max(0,\cdot)$.

In this paper, for all above feature extraction encoder, we employ CNNs as the backbone network, which effectively captures local spatial-frequency dependencies in the CSI data~\cite{yapar2023real}. Detailed architectural configurations will be specified in Sec.~\ref{subsec:architecture_training}.
Besides, we employ the angle–delay profile–based (ADP) as dissimilarity metric $d_{\text{CSI}}(\cdot,\cdot)$, 
as detailed in~\cite{stephan2024angle}, to train both the Siamese and triplet networks. 
The ADP metric quantifies the difference between two CSI samples 
by comparing their multipath structures in the joint angle–delay domain, 
thus providing a geometry-aware and physically interpretable similarity measure. For each BS $b$ with the angle–delay-domain channel
$\tilde{\bm{H}}_b$, the ADP-based dissimilarity between two CSI samples is defined as~\cite{stephan2024angle}
\begin{equation}
d_{\text{ADP}}(\mathcal{H}_i, \mathcal{H}_j)
= \sum_{b=1}^{N_{\text{BS}}}
\left(
1 -
\frac{
\langle \tilde{\bm{H}}_b(\mathcal{H}_i), \tilde{\bm{H}}_b(\mathcal{H}_j) \rangle
}{
\|\tilde{\bm{H}}_b(\mathcal{H}_i)\|_F \, 
\|\tilde{\bm{H}}_b(\mathcal{H}_j)\|_F
}
\right),
\label{eq:multiBS_ADP_dissimilarity}
\end{equation}
where $\langle \cdot, \cdot \rangle$ denotes the Frobenius inner product, 
and $\|\cdot\|_F$ represents the Frobenius norm. 
This cosine-distance–based formulation captures the degree of structural similarity 
between two multipath profiles across all BSs.

Based on the aforementioned methods, the encoder networks 
$\mathcal{G}_{\text{enc}}^{\text{ae}}(\cdot; \bm{\theta}_e)$, 
$\mathcal{G}_{\text{enc}}^{\text{siam}}(\cdot; \bm{\theta}_e)$, 
and $\mathcal{G}_{\text{enc}}^{\text{tri}}(\cdot; \bm{\theta}_e)$ 
learn to project high-dimensional CSI tensors into a compact latent space. 
Using any of these CC-based networks, the labeled dataset 
$\mathcal{D}_{\text{lab}}$ can be extended with the learned embeddings to form an enhanced fingerprint database:
\begin{align}
\mathcal{D}_{\text{lab}}^{\star} 
= \{\, (\mathcal{H}(\mathcal{L}_i), \mathcal{L}_i, \mathbf{z}_i ) 
\mid \ & \mathbf{z}_i = \mathcal{G}_{\text{enc}}^{(\cdot)}(\mathcal{H}(\mathcal{L}_i); \bm{\theta}_e),~ \nonumber \\
& i = 1, 2, \dots, N_{\text{lab}} \,\}.
\end{align}
Given a query CSI sample $\mathcal{H}_j$, 
its low-dimensional representation is obtained as 
$\mathbf{z}_j = \mathcal{G}_{\text{enc}}^{(\cdot)}(\mathcal{H}_j; \bm{\theta}_e)$.
The top-$k$ most correlated RPs are then efficiently retrieved 
from $\mathcal{D}_{\text{lab}}^{\star}$ 
through distance comparisons in the latent space, i.e.,
\begin{equation}
\mathcal{D}_{\text{ref}}(\mathcal{H}_j)
= \mathcal{G}_{\text{ret}}(\mathcal{D}_{\text{lab}}^{\star}, \mathbf{z}_j, K).
\end{equation}
The retrieval function $\mathcal{G}_{\text{ret}}$ is implemented as a $k$-nearest neighbors search in the latent space, where Euclidean distance serves as the similarity metric.
Accordingly, the overall retrieval function $\mathcal{G}(\cdot; \bm{\theta}_1)$ 
can be expressed as
\begin{equation}
\mathcal{G}(\mathcal{D}_{\text{lab}},\mathcal{H}_j; {\bm{\theta}_1})
= \mathcal{G}_{\text{ret}}\big(\mathcal{D}_{\text{lab}}^{\star}, \mathcal{G}_{\text{enc}}^{(\cdot)}(\mathcal{H}_j; {\bm{\theta}_e}),~K
\big).
\end{equation}
Note that $\bm{\theta}_1$ is equivalent to $\bm{\theta}_e$, as the retrieval function $\mathcal{G}_{\text{ret}}$ is a non-parametric operation that relies solely on the latent representations generated by the encoder $\mathcal{G}_{\text{enc}}^{(\cdot)}(\cdot; \bm{\theta}_e)$.
This channel charting–based representation and retrieval mechanism 
provide an interpretable, low-complexity, and structure-aware foundation 
for the subsequent localization network. 
However, it is worth noting that the aforementioned CC-based methods do not guarantee globally optimal geometric consistency. 
Specifically, the learned representations in CC inherently reflect the dissimilarity metric employed during training, but this metric often fails to maintain consistent correlation with physical distance (e.g., ADP dissimilarity exhibits degraded distance correlation when multipath characteristics change sharply across propagation boundaries).
Despite this, by selecting RPs that exhibit high similarity 
to the query CSI in the latent space, 
these methods are able to capture locally correlated fingerprints 
that serve as effective and informative references for localization. 
The design of more advanced channel charting techniques 
that can further enhance performance and robustness 
remains an open problem for future research.

\subsection{RP-Correlation-Aware GAT-based Localization}
\label{subsect:GAT}
To implement the localization network $\mathcal{F}(\cdot\,;\bm{\theta}_2)$ defined in \eqref{eqn:loc_func} of Problem~1, we first construct a correlation graph between the query and its retrieved RPs, and then process it with a GAT.
Based on the retrieved top-$k$ RPs $\mathcal{D}_{\text{ref}}(\mathcal{H}_j)$ corresponding to the query CSI sample $\mathcal{H}_j$,
a graph structure is constructed to model the spatial and semantic correlations between the query and its retrieved RPs.
In the following, we introduce the proposed GAT-based localization method, 
which leverages this graph representation to perform correlation-aware feature aggregation and position estimation.

\subsubsection{Graph Construction}
In the proposed framework, the GAT operates on a graph structure constructed from the retrieved reference set. 
Specifically, we define a graph $G = (\mathcal{V}, \mathcal{A})$ for each query CSI sample $\mathcal{H}_j$, where:

\begin{itemize}
    \item \textbf{Nodes ($\mathcal{V}$):}  
    Since the proposed framework constructs the graph based on the top-$K$ retrieved RPs, the resulting graph contains a total of $K+1$ nodes, i.e., $\mathcal{V} = \{v_k \mid k = 0, 1, \dots, K\}$, where $v_0$ corresponds to the query CSI sample and the remaining $K$ nodes correspond to its retrieved RPs. Each node $v_k$ is associated with an initial node feature $\mathbf{x}_k^{(-1)}$, which jointly encodes the CSI representation and the corresponding spatial information. Specifically, for the retrieved RPs, the node feature is defined as $\mathbf{x}_k^{(-1)} = \{ \mathcal{H}_k,\mathcal{L}_k \}, \quad k = 1, \dots, K$, where $\mathcal{H}_k$ denotes the CSI tensor of the $k$-th retrieved RP, $\mathcal{L}_k$ denotes its known position. For the query node $v_0$, since its true location is unknown during inference, we define its initial node feature as $\mathbf{x}_0^{(-1)} = \{ \mathcal{H}_0, \textbf{0} \}$, where $\mathbf{0}$ denotes a zero vector with the same dimensionality as $\mathcal{L}_k$. This placeholder design enforces consistency between training and inference phases by maintaining identical feature structures even when ground-truth positions are accessible during training, thereby preventing premature exploitation of label information.

    \item \textbf{Edges ($\mathcal{A}$):}  
    To capture potential correlations among the nodes, we construct a fully connected graph, 
    where each edge $A_{pq}$ represents the similarity between two CSI tensors $\mathcal{H}_p$ and $\mathcal{H}_q$.
    The edge weight is initialized as the ADP-based similarity score $1 / d_{\text{ADP}}(\mathcal{H}_i, \mathcal{H}_j)$, which provide geometric priors that inform the GAT's attention mechanism during training.
    
\end{itemize}

Once the graph $G = (\mathcal{V}, \mathcal{A})$ is constructed, 
the localization task is formulated as learning to infer the position of the query sample $\mathcal{H}_j$ 
by aggregating information from its correlated RPs through GAT. 

\subsubsection{GAT-based Localization}

Each node $v_k$ is initialized with a feature vector 
$\mathbf{x}_k^{(0)} = \Phi(\mathbf{x}_k^{(-1)})$, 
where $\Phi(\cdot)$ denotes a feature extraction function.
As mentioned before, $\mathbf{x}_k^{(-1)}$ incorporates both CSI and spatial information, which we process using a CNN and MLP respectively, aggregating their outputs to form the initial node representation. 
The detailed architecture will be discussed in Sec.~\ref{subsec:architecture_training}.

After obtaining the node features $\{\mathbf{x}_k^{(0)}\}$ from the feature extraction module, 
we employ the GAT to adaptively aggregate relational information among all nodes 
and capture the underlying correlations between the query CSI and its retrieved RPs. At each GAT layer ($l\geq0$), node features are updated through an attention-based aggregation mechanism.
For the $l$-th layer, each node $v_k$ computes attention scores $e_{kq}^{(l)}$ for all other nodes $v_q \in \mathcal{V} \setminus \{k\}$ through a unified transformation:
\begin{equation}
e_{kq}^{(l)} = \psi\!\left( \mathbf{a}^{(l)\top} \left[ \mathbf{W}^{(l)} \mathbf{x}_k^{(l)} \Vert \mathbf{W}^{(l)} \mathbf{x}_q^{(l)} \right] \right)
\label{eq:gat_attention_score}
\end{equation}
where $\mathbf{W}^{(l)}$ is a shared learnable weight matrix applying linear transformation to node features, $\Vert$ denotes feature concatenation, $\mathbf{a}^{(l)}$ is a learnable attention vector, $\psi(\cdot)$ introduces nonlinearity via activation function \cite{velickovic2017graph}.
The attention coefficient $\alpha_{kq}^{(l)}$, which quantifies the relative importance of node $v_q$ to node $v_k$, is obtained by normalizing the scores across all neighboring nodes using the softmax function:
\begin{equation}
\alpha_{kq}^{(l)} = \frac{\exp\!\left(e_{kq}^{(l)}\right)}{\sum_{r \in \mathcal{V} \setminus \{k\}} \exp\!\left(e_{kr}^{(l)}\right)}.
\label{eq:GAT_attention_coeff}
\end{equation}

Both $\mathbf{W}^{(l)}$ and $\mathbf{a}^{(l)}$ are optimized during training through backpropagation, enabling the model to automatically discover significant correlations between nodes.
The updated feature representation of node $v_k$ is then computed as \cite{velickovic2017graph}:
\begin{equation}
\mathbf{x}_k^{(l+1)} = 
\psi\!\left(\sum_{q \in \mathcal{V} \setminus \{k\}} 
\alpha_{kq}^{(l)} \mathbf{W}^{(l)} \mathbf{x}_q^{(l)} \right).
\label{eq:GAT_update}
\end{equation}
To further enhance model robustness and expressiveness, 
multi-head attention is adopted, 
where multiple independent attention mechanisms are applied in parallel, 
and their outputs are concatenated.

After propagating through $N_{\text{GAT}}$ layers, 
the final node representations $\{\mathbf{x}_k^{(N_{\text{GAT}})}\}$ are obtained, 
which encode rich contextual and spatially correlated information from the graph structure. 
To predict the position of the query CSI sample $\mathcal{H}_j$, 
a localization head $\mathcal{F}_{\text{loc}}(\cdot)$ is employed to project 
the final feature representation of the query node $v_0$ 
onto the coordinate space:
\begin{equation}
\widehat{\mathcal{L}}_j = 
\mathcal{F}_{\text{loc}}\!\left(\mathbf{x}_0^{(N_{\text{GAT}})}\right),
\label{eq:localization_head}
\end{equation}
where $\mathcal{F}_{\text{loc}}(\cdot)$ is implemented as a series of MLP layers that perform nonlinear regression from 
the learned latent representation to the physical coordinate domain.
To train the localization model, we minimize the mean squared error (MSE) between predicted and ground-truth positions:
\begin{equation}
\mathcal{J}_{\text{loc}} = \frac{1}{N_{\text{lab}}} \sum_{j=1}^{N_{\text{lab}}} \left\| \widehat{\mathcal{L}}_j - \mathcal{L}_j \right\|^2_2,
\label{eq:loss_loc}
\end{equation}
where $N_{\text{lab}}$ denotes the number of training samples.
The overall set of learnable parameters for the localization mapping function 
$\mathcal{F}(\cdot;{\bm{\theta}_2})$ includes all parameters 
of the feature extraction module $\Phi(\cdot)$, 
the GAT layers, and the final localization head $\mathcal{F}_{\text{loc}}(\cdot)$. 
Together, these components enable the model to adaptively integrate 
both the query CSI and the retrieved RPs 
for accurate and correlation-aware position estimation.

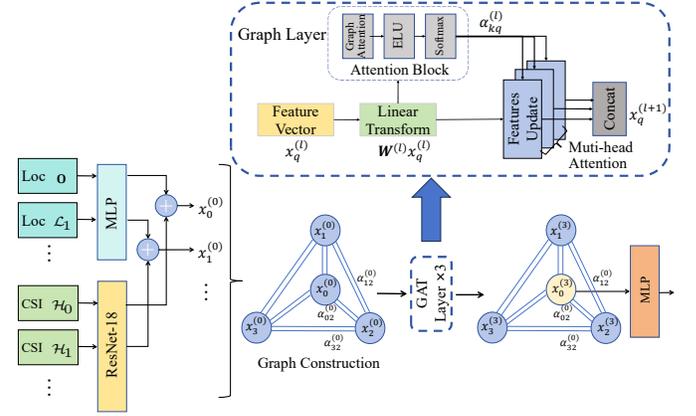
\begin{figure}[t]
    \centering
    \begin{tikzpicture}
    \node (image) [anchor=south west]{\includegraphics[width=1.0\linewidth]{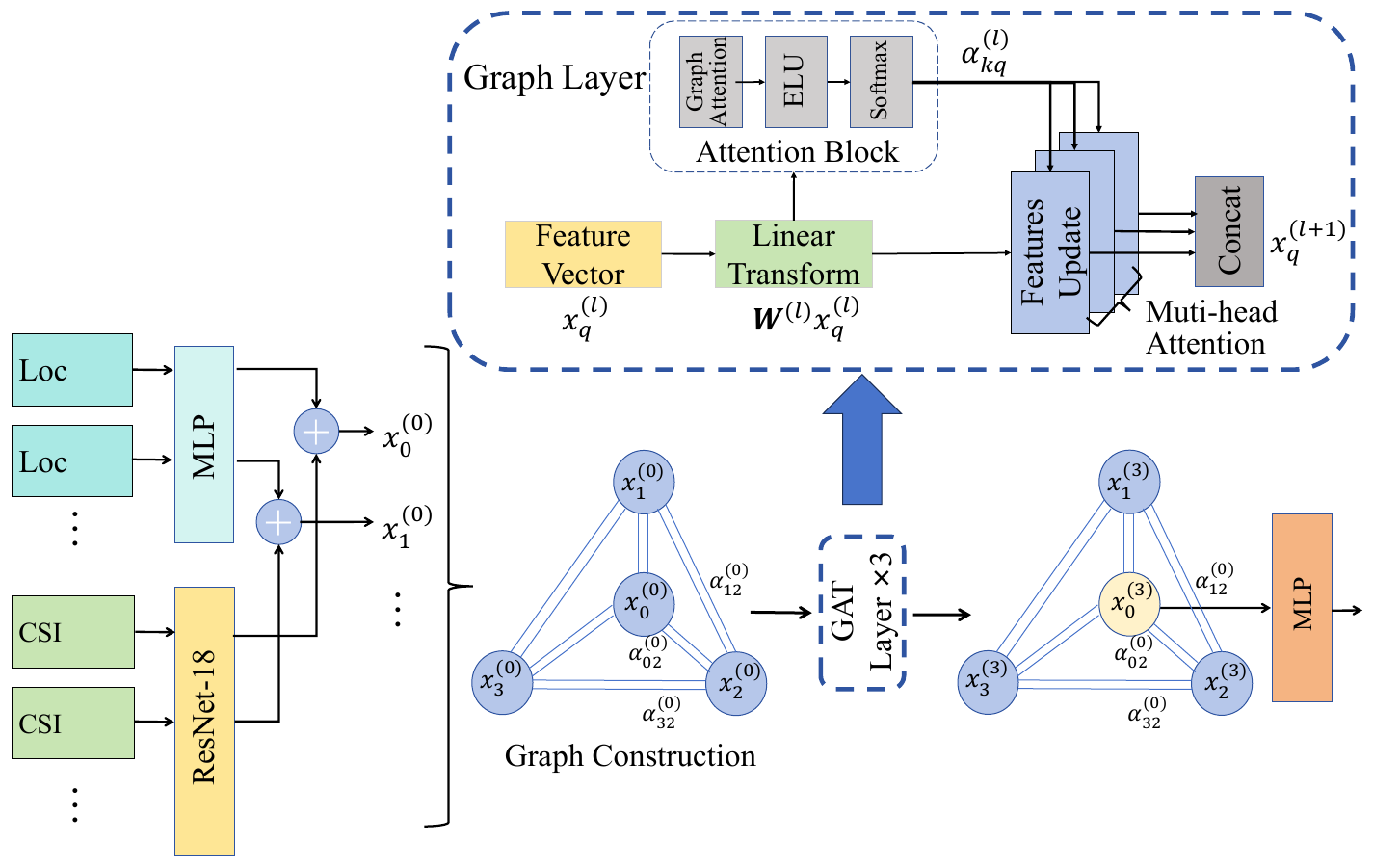}};
    \gettikzxy{(image.north east)}{\ix}{\iy};

    \node at (0.080*\ix,0.308*\iy)[rotate=0,anchor=north]{{\fontsize{4pt}{4.8pt} $\mathcal{H\!}_0$}};
    \node at (0.080*\ix,0.21*\iy)[rotate=0,anchor=north]{{\fontsize{4pt}{4.8pt} $\mathcal{H\!}_1$}};

    \node at (0.078*\ix,0.595*\iy)[rotate=0,anchor=north]{{\fontsize{4pt}{4.8pt} $\bf{0}$}};
    \node at (0.080*\ix,0.50*\iy)[rotate=0,anchor=north]{{\fontsize{4pt}{4.8pt} $\mathcal{L\!}_1$}};
    \end{tikzpicture}
    \caption{GAT-based neural network architecture for localization module.}
    
    \label{fig:framework_loc}
    \vspace{-3mm}
\end{figure}

\subsection{Neural Network Architecture and Training}
\label{subsec:architecture_training}

In this subsection, we present the detailed neural network architectures adopted in the proposed CC-GAT framework, along with the corresponding training configurations. 

\subsubsection{Architecture for CC}
The multi-BS channel tensor $\mathcal{H}$ encapsulates rich structural information 
across the spatial (BS), frequency (subcarrier), and temporal (delay) domains. 
To effectively capture these spatial–frequency correlations, 
we adopt a CNN-based encoder as the backbone 
for the channel charting module. Specifically, we employ a ResNet-18 \cite{he2016deep} to extract features 
from the input CSI tensor $\mathcal{H}$. 
The output feature map of the ResNet-18 is flattened and subsequently passed through 
two MLP layers with output dimensions of 128 and $d$, respectively. 
The resulting 2-dimensional representation vector serves as the learned channel chart representation $\mathbf{z} \in \mathbb{R}^2$. Note that the channel chart representation $\mathbf{z}$ can be generalized to higher dimensions but is set to $\mathbb{R}^2$
in experiments for geometric visualization. 
For the autoencoder, a symmetric decoder network mirroring the ResNet-18 encoder is employed to reconstruct the original CSI tensor, 
and the entire network is trained by minimizing the reconstruction loss defined in~\eqref{eq:ae_loss}. 
For the Siamese and triplet-based CC models, the same ResNet-18 encoder is utilized 
to generate embeddings for each CSI input, 
which are then trained using the ADP-dissimilarity-based loss~\eqref{eq:cc_Siamese} and ~\eqref{eq:triplet_loss}, respectively.

\begin{figure}[th]
\centering
\includegraphics[width = 2.5in, height=1.8in]{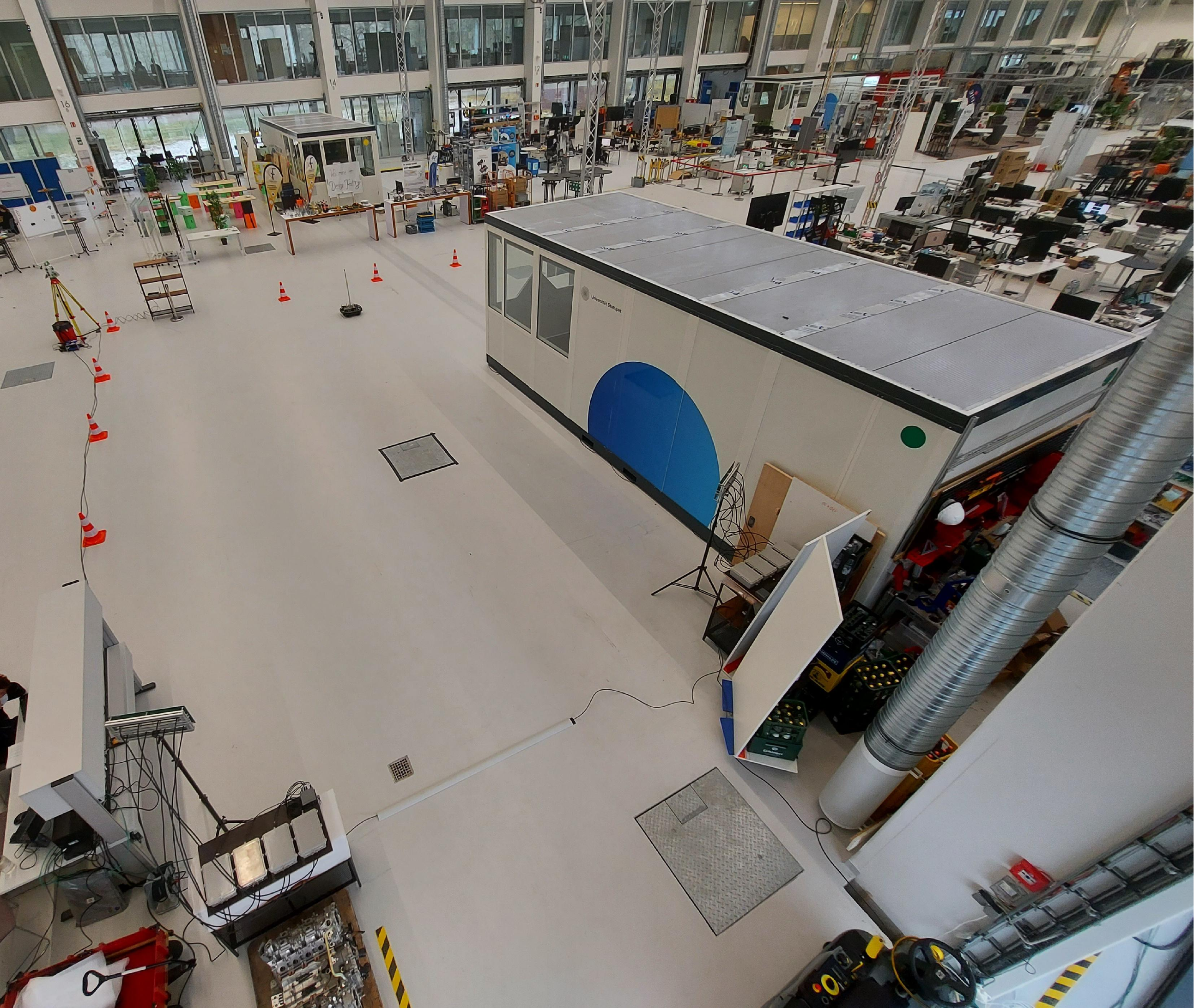}
\caption{A photographic depiction of the environment where the DICHASUS dataset was measured\cite{euchner2021distributed}.}
\label{fig:DICHASUS}
\end{figure}

\begin{figure}[th]
\centering
\includegraphics[width = 2.5in, height=1.8in]{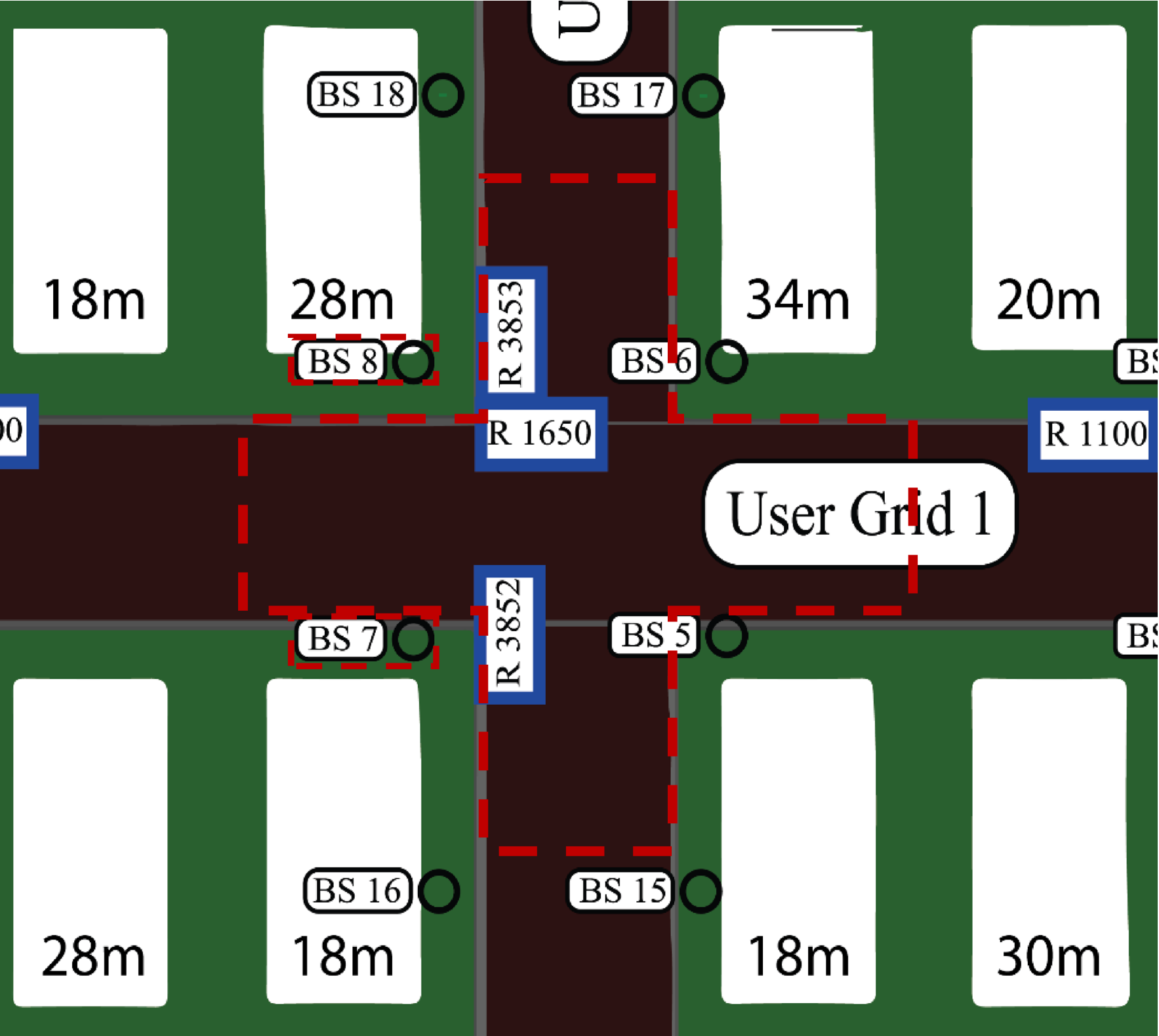}
\caption{Top-view layout of the DeepMIMO dataset scenario, highlighting the sampling area and the selected base stations\cite{alkhateeb2019deepmimo}.}
\label{fig:DEEPMIMO}
\end{figure}

\subsubsection{Architecture for Localization}

The overall network architecture for the proposed localization module is illustrated in Fig.~\ref{fig:framework_loc}. 
For each query CSI sample $\mathcal{H}_0$ and its retrieved reference set 
$\{\mathcal{H}_1, \dots, \mathcal{H}_K\}$, 
we first extract CSI feature representations using a ResNet-18 backbone~\cite{he2016deep}, which shares the identical architecture with the CC encoder in Sec. \ref{subsec:architecture_training}. Denoted as $\Phi_1(\mathcal{H}_k)$ for $k = 0, \dots, K$, this feature extractor outputs 256-dimensional representations. 
Meanwhile, a one-layer MLP is employed to encode the location information associated with each RP, denoted as $\Phi_2(\mathcal{L}_k)$ for $k = 0, \dots, K$, which projects the $\mathcal{L}_k$ to the same dimension as the CSI features.
The CSI features and the corresponding location embeddings are then combined through element-wise addition to form the initial node representation, i.e., $\mathbf{x}_k^{(0)} = \Phi(\mathbf{x}_k^{(-1)}) = \Phi_1(\mathcal{H}_k) + \Phi_2(\mathcal{L}_k)
$.
These extracted feature embeddings are then utilized, together with 
the ADP-based similarity scores 
$1 / d_{\text{ADP}}(\mathcal{H}_i, \mathcal{H}_j)$, 
to construct a fully connected graph structure as described in Section~\ref{subsect:GAT}. Subsequently, a three-layer GAT 
is employed to perform adaptive feature aggregation 
across all nodes within the constructed graph, 
capturing the correlation structure among the query and reference CSI samples. 
After the GAT module, the final node representation corresponding to 
the query CSI $\mathcal{H}_0$ is fed into a MLP head 
to regress the estimated position, and the localization network is trained by minimizing the MSE loss defined in~\eqref{eq:loss_loc}. 
For all GAT layers, we employ the exponential linear unit (ELU) activation function \cite{Clevert2015}.
The parameter configurations of the GAT and MLP modules for localization are summarized in Table~\ref{tab:gat_params}.

\begin{table}[tbp]
  \centering
  \caption{GAT Model Architecture and Hyperparameters}
  \label{tab:gat_params}
  \begin{tabular}{@{}lcccc@{}}
    \toprule
    \textbf{Layer} & \textbf{Input} & \textbf{Output} & \textbf{Heads} & \textbf{Activation} \\
    \midrule
    GAT Layer 1 & 256 & 512 & 4 & ELU \cite{Clevert2015}\\
    GAT Layer 2 & 512 & 2048 & 8 & ELU \\
    GAT Layer 3 & 2048 & 64 & 1 & ELU \\
    \midrule
     MLP Layer & 64 & 2 & - & - \\
    \bottomrule
  \end{tabular}
\end{table}

\begin{table}[h]
\centering
\caption{Dataset configurations for experimental evaluation.}
\label{tab:dataset_config}
\begin{tabular}{p{3cm} cc}
\toprule
Parameter & DICHASUS & DeepMIMO \\
\midrule
$N_{\text{BS}}$ & 4 & 2(BS 7, 8)  \\
$N_r$ & 8  & 32  \\
$N_{\text{SC}}$ & 1024 & 128 \\
Carrier frequency & 1.272 GHz & 28 GHz \\
Bandwidth & 50 MHz & 50 MHz \\
Total samples & \multicolumn{2}{c}{10,000} \\
\bottomrule
\end{tabular}
\label{table:setting}
\end{table}

The training protocol follows a two-stage paradigm to ensure geometric consistency and task specialization. In the first stage, the CC module is trained in a completely self-supervised manner using the entire dataset. Subsequently, the localization module undergoes supervised training exclusively on labeled fingerprint data $\mathcal{D}_{\text{lab}}$. Crucially, during this second stage, all parameters of the pre-trained CC encoder remain frozen. Only the GAT layers and regression head are updated through backpropagation of the MSE loss. This decoupled optimization strategy prevents overfitting while leveraging the geometrically meaningful embeddings learned during CC.

\section{Experimental Results} \label{sect:Experiment}
\subsection{Experiment Settings}
\label{subsec:settings}

\subsubsection{Datasets and Configurations}

For our experimental evaluation, we employ two distinct datasets to validate the proposed architecture: a real-world indoor measurement dataset DICHASUS \cite{euchner2021distributed} and a ray-tracing-based outdoor simulation dataset DeepMIMO \cite{alkhateeb2019deepmimo}.

\begin{table*}[htbp]
\centering
\caption{MAE(m) comparison of various localization methods under different labeled dataset sizes for training $N_{\text{lab}}$ on DICHASUS and DeepMIMO scenarios.}
\label{tab:mae_comparison_combined}
\small
\begin{tabular}{@{}m{2cm} m{1.2cm} @{\hspace{5em}} cccc p{0.2cm} cccc@{}}
\toprule
\multicolumn{2}{c}{\textbf{Scene}} & \multicolumn{4}{c}{\textbf{DICHASUS}} & & \multicolumn{4}{c}{\textbf{DeepMIMO}} \\
\midrule
\multicolumn{2}{c}{\textbf{Labeled Number $N_{\text{lab}}$}} & \textbf{100} & \textbf{500} & \textbf{1000} & \textbf{2000} & & \textbf{100} & \textbf{500} & \textbf{1000} & \textbf{2000} \\
\midrule
\multicolumn{2}{c}{\textbf{Baseline Method}} \\
\midrule
\multicolumn{2}{c}{WKNN} & 4.42 & 4.11 & 3.87 & 3.66 & & 25.88 & 18.15 & 14.01 & 10.92 \\
\multicolumn{2}{c}{CNN-based} & 3.61 & 2.23 & 1.62 & 1.31 & & 12.89 & 9.62 & 7.71 & 5.44 \\
\multicolumn{2}{c}{Transformer-based} & 3.83 & 2.72 & 2.03 & 1.65 & & 17.57 & 13.26 & 10.28 & 8.02 \\
\midrule
\multicolumn{2}{c}{\textbf{Proposed Retrieval-Assisted Framework}}\\
\midrule
\multirow{2}{*}{\centering Autoencoder} & \centering GAT & 2.27 & 1.44 & 1.12 & 0.87 & & 9.47 & 7.14 & 4.40 & 2.93 \\
 & \centering CNN & 3.24 & 1.77 & 1.21 & 0.92 & & 9.68 & 8.27 & 5.77 & 3.42 \\
 & \centering Transformer & 2.27 & 1.92 & 1.60 & 1.23 & & 10.82 & 9.03 & 6.54 & 4.38 \\
\midrule
\multirow{2}{*}{\centering Siamese} & \centering GAT & \textbf{1.53} & \textbf{0.96} & \textbf{0.80} & \textbf{0.71} & & \textbf{8.53} & \textbf{6.49} & \textbf{3.56} & \textbf{2.57} \\
 & \centering CNN & 1.69 & 1.11 & 0.93 & 0.83 & & 8.56 & 7.08 & 4.15 & 2.79 \\
 & \centering Transformer & 1.78 & 1.49 & 1.32 & 1.04 & & 9.25 & 7.69 & 4.92 & 3.67 \\
\midrule
\multirow{2}{*}{\centering Triplet} & \centering GAT & 1.53 & 1.13 & 0.88 & 0.76 & & 8.75 & 6.63 & 3.86 & 2.75 \\
 & \centering CNN & 1.56 & 1.16 & 0.96 & 0.89 & & 8.77 & 7.45 & 4.84 & 2.88 \\
 & \centering Transformer & 1.83 & 1.57 & 1.35 & 1.08 & & 9.56 & 7.80 & 5.14 & 3.72 \\
\midrule
\multirow{2}{*}{\centering Isomap} & \centering GAT & 1.55 & 1.30 & 0.93 & 0.85 & & 8.41 & 6.98 & 3.92 & 2.83 \\
 & \centering CNN & 1.89 & 1.35 & 1.08 & 0.92 & & 8.72 & 7.61 & 5.74 & 3.06 \\
 & \centering Transformer & 1.99 & 1.74 & 1.41 & 1.15 & & 9.71 & 8.09 & 5.38 & 4.03 \\
\midrule
\multicolumn{2}{c}{\textbf{Ideal Performance}}\\
\midrule
\multirow{2}{*}{\centering Real Location} & \centering GAT & 0.86 & 0.59 & 0.45 & 0.38 & & 6.62 & 4.28 & 2.78 & 1.50 \\
 & \centering CNN & 1.13 & 0.76 & 0.64 & 0.55 & & 7.38 & 4.80 & 2.83 & 2.16 \\
 & \centering Transformer & 1.59 & 1.28 & 1.10 & 0.85 & & 8.24 & 5.33 & 3.55 & 2.47 \\
\midrule
\multirow{2}{*}{\centering ADP} & \centering GAT & 1.49 & 0.89 & 0.74 & 0.65 & & 8.38 & 6.47 & 3.25 & 2.38 \\
 & \centering CNN & 1.55 & 1.05 & 0.85 & 0.74 & & 8.43 & 6.82 & 3.84 & 2.57 \\
 & \centering Transformer & 1.74 & 1.43 & 1.20 & 0.98 & & 9.01 & 7.33 & 4.56 & 3.34 \\
\bottomrule
\end{tabular}
\end{table*}

We selected the dichasus-cf0x \cite{dataset-dichasus-cf0x} subset from the DICHASUS system, which measures the propagation channel, with the environmental context illustrated in Fig.~\ref{fig:DICHASUS}. As detailed in \cite{dataset-dichasus-cf0x}, the measurement was conducted in an industrial indoor environment of approximately 14 m × 14 m in an L-shaped configuration. The setup involved $N_{\text{BS}}=4$ separate antenna arrays, each equipped with $N_r=8$ antennas. A total of $N_{\text{SC}}=1024$ OFDM channel coefficients were measured at a carrier frequency of 1.272 GHz with a 50 MHz channel bandwidth. The single dipole transmit antenna is mounted on a robot traversing predefined trajectories within the area.

For the DeepMIMO dataset, we adopt the O1 scenario as described in \cite{alkhateeb2019deepmimo}, which models a typical urban outdoor environment with intersecting streets. The environmental layout is illustrated in Fig.~\ref{fig:DEEPMIMO}, where the sampling area is centered on the cross-junction region, and we specifically activate $N_{\text{BS}}=2$ BSs (i.e. BS 7 and 8) for data collection. In this setup, each UE is equipped with a single antenna, while each BS employs a ULA with $N_r=32$ antennas. The system operates at a carrier frequency of 28 GHz with a bandwidth of 50 MHz, and CSI is captured over $N_{\text{SC}}=128$ OFDM subcarriers. All experiments assume two-dimensional  positioning as both DICHASUS and DeepMIMO datasets provide user positions on a horizontal plane. 

For each dataset, we collect a total of 10,000 channel samples and only a subset of these samples has known true positions. This labeled subset serves as a reference database storing CSI fingerprints associated with known positions for supervised learning, while the remaining samples constitute the test set, and we adopt this configuration to simulate few-shot scenarios with limited labeled data. The settings for the experimental evaluation are summarized in Table~\ref{table:setting}.

\subsubsection{Baseline Methods}

To demonstrate the effectiveness of our proposed retrieval-assisted fingerprinting localization framework, we evaluated a few types of typical localization algorithms mentioned in Section.~\ref{subsect:prelim}.
\begin{itemize}
    \item \textbf{WKNN \cite{zhou2021integrated} fingerprinting:} A typical similarity-based localization method with the hyperparameter K set to 20 in our evaluation, 
    where we employ the ADP dissimilarity as the distance measure.

    \item \textbf{CNN-based \cite{GAO_2024CNN} fingerprinting:} A learning-based method where the localization network $\mathcal{F}(\cdot)$, as defined in Eq.~\eqref{eq:learning_loss}, is instantiated as a CNN, specifically ResNet18.

    \item \textbf{Transformer-based \cite{vaswani2017attention} fingerprinting:} A learning-based method  with $\mathcal{F}(\cdot)$ set to Transformer, with specific model architecture and hyperparameter settings detailed in \cite{gong2023deep}.
\end{itemize}

\subsubsection{Proposed Framework Components}

To validate the rationale for using CC as $\mathcal{G}(\cdot\,;\bm{\theta}_1)$ and GAT as $\mathcal{F}(\cdot\,;\bm{\theta}_2)$ in our framework, we perform an extensive evaluation of various combinations of methods. For $\mathcal{G}(\cdot\,;\bm{\theta}_1)$, we consider the following approaches:
\begin{itemize}
    \item \textbf{Autoencoder, Siamese Network and Triplet-based Method:} The three principal self-supervised learning CC algorithms are detailed in Section.~\ref{subsect:CC}

    \item \textbf{Isomap\cite{tenenbaum2000global}:} A traditional manifold learning method that serves as a non-parametric CC solution. However, it cannot generalize to unseen CSI samples during inference, limiting its practical applicability.

    \item \textbf{Real Location:} Select RPs based on the true physical distance to the query location. This method represents the theoretical upper bound of the framework since it leverages perfect knowledge to select optimally correlated RPs. It is infeasible in practice because the true location of the query CSI is unknown during inference.

    \item \textbf{ADP Dissimilarity:} Directly utilizes the ADP dissimilarity metric as the retrieval criterion. Since ADP serves as the training target for learning-based CC, this approach represents the performance upper bound, as it bypasses approximation errors inherent in learned embeddings.
    
\end{itemize}
For $\mathcal{F}(\cdot\,;\bm{\theta}_2)$, we consider the following approaches:
\begin{itemize}
    \item \textbf{GAT:} The proposed method trained with RPs of graph-structured data.

    \item \textbf{CNN:} A CNN model based on ResNet18. Since CNN cannot accept graph-structured data, we stack all the features along the RP dimension to serve as input.

    \item \textbf{Transformer:}A Transformer-based model with the same configuration as \cite{gong2023deep}, employing the same input processing method as the CNN approach.
    
\end{itemize}

We perform a comprehensive evaluation of all the methods considered, encompassing the baselines and various combinations of components of the proposed retrieval-assisted framework. The overall experimental results are presented in Table~\ref{tab:mae_comparison_combined}. For the results presented in Table~\ref{tab:mae_comparison_combined}, we fix the number of top-$K$ most correlated RPs $N_{\text{ref}}$ to 20. The evaluation spans both DICHASUS and DeepMIMO scenes, assessing performance across a range of labeled dataset sizes, from $N_{\text{lab}}=100$ to $N_{\text{lab}}=2000$, respectively.

\subsection{Performance Comparison with Baseline Methods}
\label{subsec:Baseline Methods}

\begin{figure}
\centering
\includegraphics[width = 3.5in, height=2.5in]{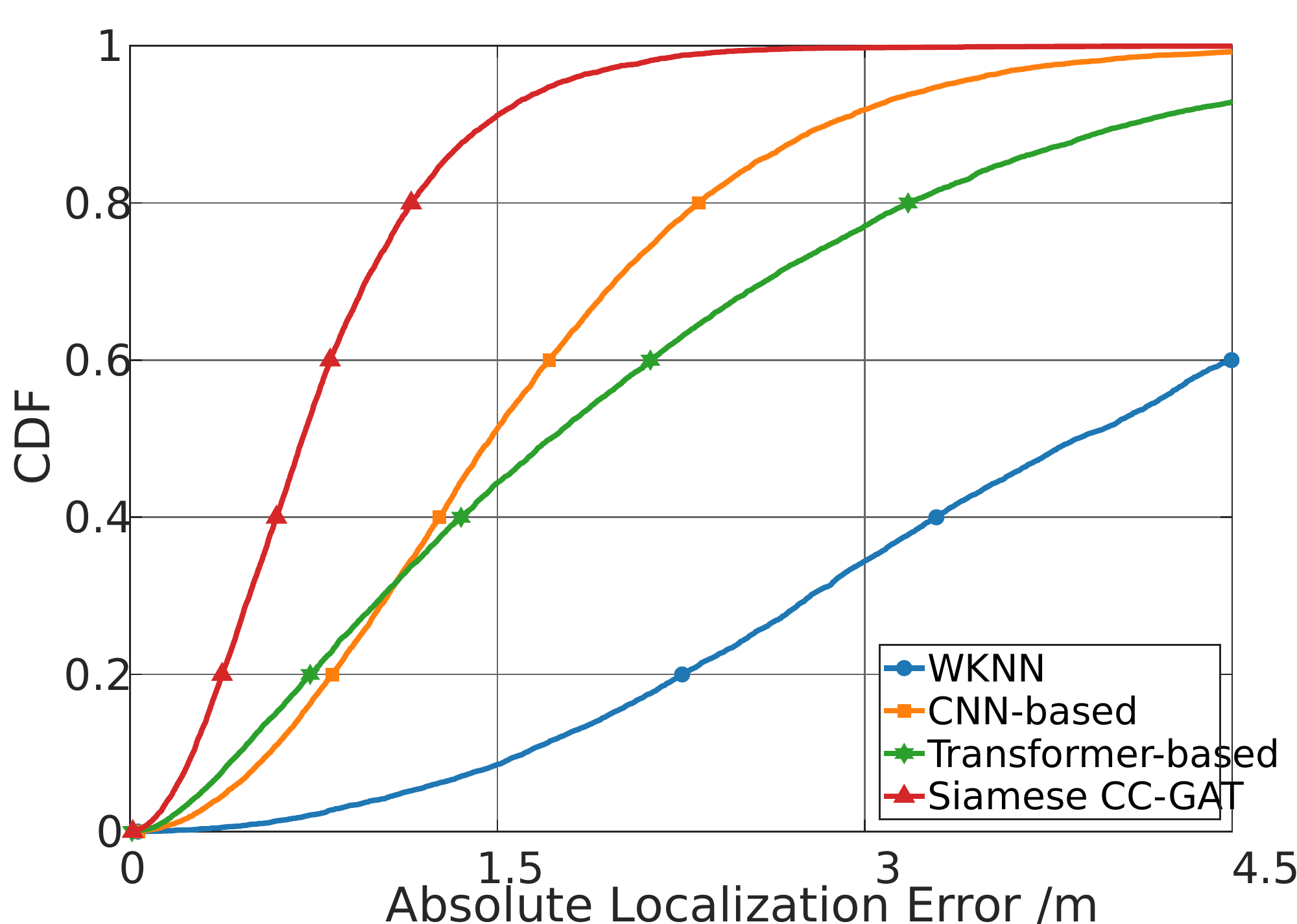}
\caption{The localization error CDF between the proposed framework and baselines under the scenario of $N_{\text{lab}}$ = 1000 on the DICHASUS scene.}
\label{fig:CDF_baseline}
\end{figure}

To comprehensively evaluate the effectiveness of our proposed retrieval-assisted fingerprinting framework, we first compare the performance against all baseline methods. Fig.~\ref{fig:CDF_baseline} presents the cumulative distribution function (CDF) of localization errors for different approaches in the DICHASUS dataset with $N_{\text{lab}} = 1000$ and $N_{\text{ref}} = 20$.
The proposed GAT-Siamese CC method demonstrates outstanding localization accuracy, achieving 50\% and 90\% localization errors of 0.70~m and 1.45~m, respectively. As shown in Fig.~\ref{fig:CDF_baseline}, our method significantly outperforms all baseline approaches, reducing MAE by 50.6\% compared to CNN-based, 60.5\% compared to Transformer-based, and 79.3\% compared to WKNN.

Although Transformer architectures have shown state-of-the-art performance in data-rich domains, their effectiveness is limited due to the substantial data requirements for learning complex intrinsic relationships within channel characteristics. Therefore, the performance of the Transformer-based baseline is relatively limited.
Our method demonstrates comprehensive superiority over all baseline methods in different sizes of labeled dataset $N_{\text{lab}}$. Specifically, in extreme low-data regimes such as $N_{\text{lab}} = 100$, our method achieves 1.53~m MAE, representing 57.6\% and 60.1\% improvements over CNN-based and Transformer-based approaches, respectively. This demonstrates exceptional generalization capability when labeled data are scarce.

The same trend is observed in the DeepMIMO scene, though with generally higher absolute localization errors due to the more complex propagation environment characterized by richer multipath effects and larger spatial variations. Despite these challenges, our framework maintains consistent performance advantages over all baseline methods, reducing MAE by 53.8\%, 65.3\%, and 74.5\% compared to CNN-based, Transformer-based, and WKNN, respectively, which validates its robustness across diverse wireless environments.

\subsection{Performance Comparison with Different RP Retrieval Mechanisms}

\begin{figure}
\centering
\includegraphics[width = 3.5in, height=2.5in]{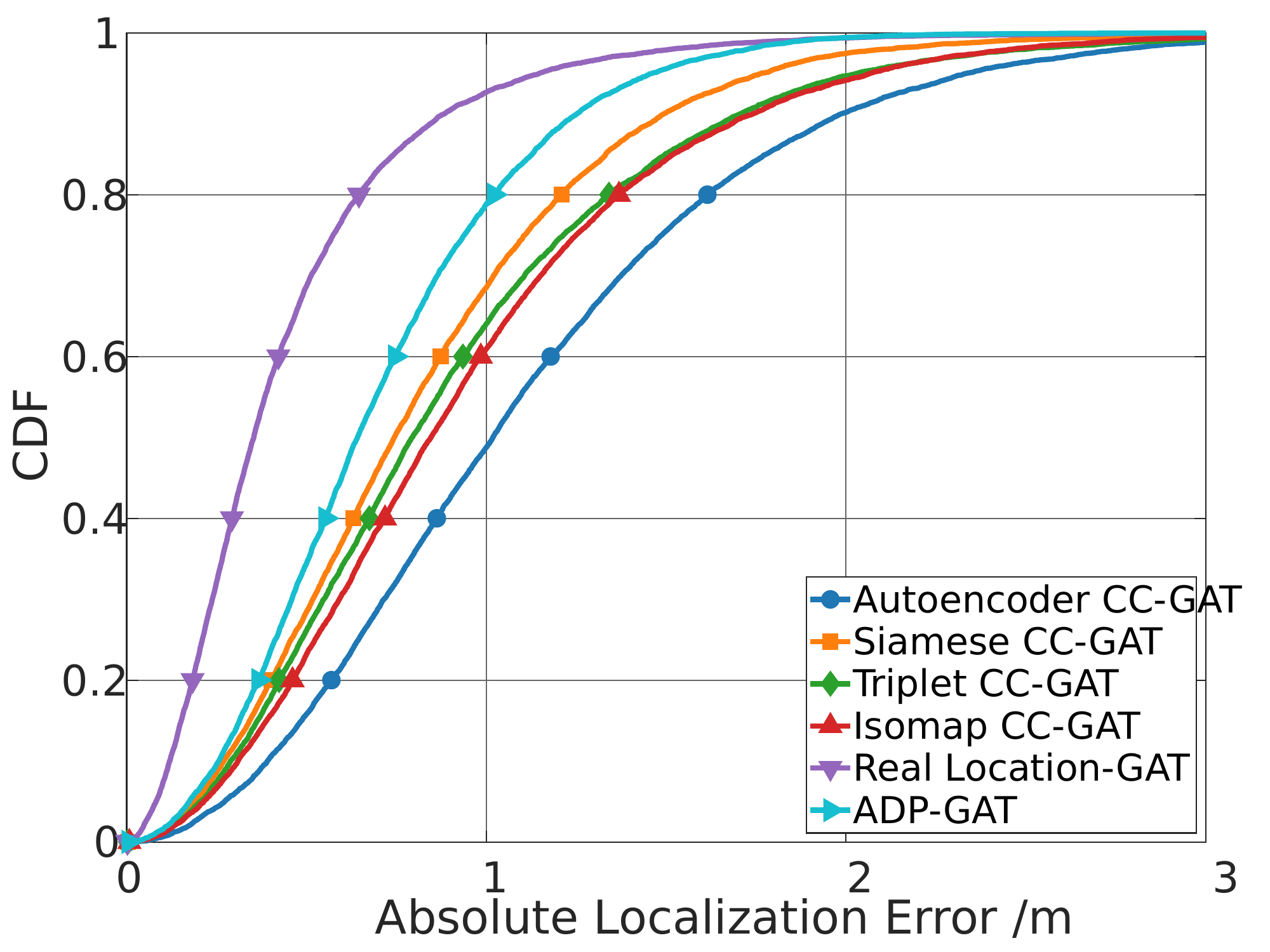}
\caption{The localization error CDF with different RP selection strategies using GAT as $\mathcal{F}(\cdot\,;\bm{\theta}_2)$ under the scenario of $N_{\text{lab}}$ = 1000 on the DICHASUS scene.}
\label{fig:CDF_CC}
\end{figure}

To evaluate the effectiveness of different RP retrieval mechanisms within our proposed framework, we first analyze the localization performance when employing GAT as the localization function $\mathcal{F}(\cdot\,;\bm{\theta}_2)$ with various RP selection methods $\mathcal{G}(\cdot\,;\bm{\theta}_1)$. Fig.~\ref{fig:CDF_CC} presents the cumulative distribution of localization errors for different RP selection strategies on the DICHASUS dataset with $N_{\text{lab}}=1000$.

The choice of channel charting method for learning low-dimensional embeddings significantly affects localization accuracy. Among the practical CC-based approaches, the Siamese network achieves the best performance, closely followed by the Triplet method and the traditional manifold learning method Isomap, while the Autoencoder demonstrates relatively weaker results. The superior performance of similarity-based learning methods over the reconstruction-based Autoencoder validates the advantage of explicitly preserving relative similarity relationships for localization tasks. 
These performance trends are consistently observed for different dataset sizes labeled in Table~\ref{tab:mae_comparison_combined}.

\begin{figure*}[htbp]
    \centering
    \begin{subfigure}{0.19\textwidth}
        \includegraphics[width=\linewidth]{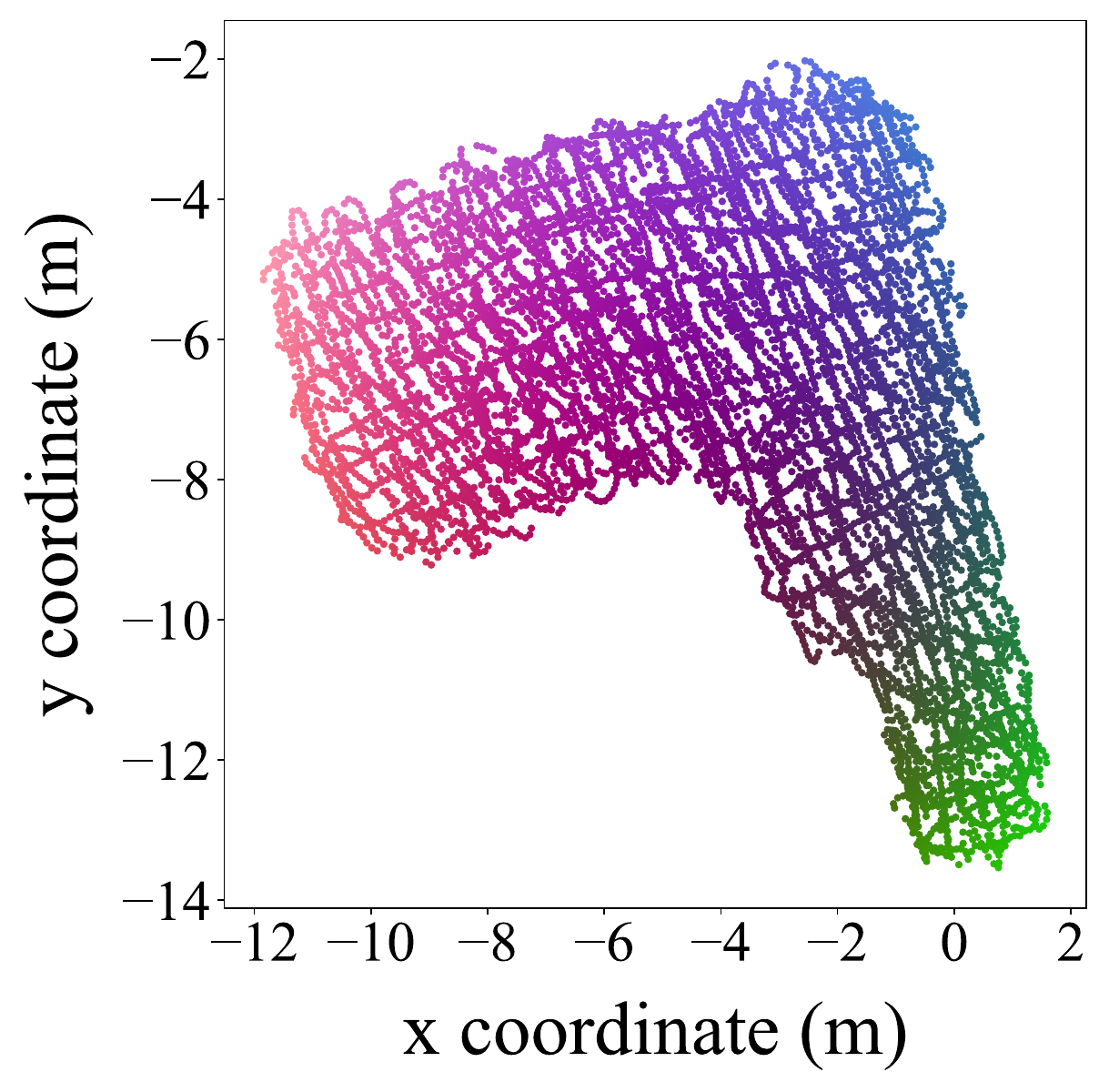}
        \caption{}
        \label{fig:top-left}
    \end{subfigure}
    \hfill
    \begin{subfigure}{0.19\textwidth}
        \includegraphics[width=\linewidth]{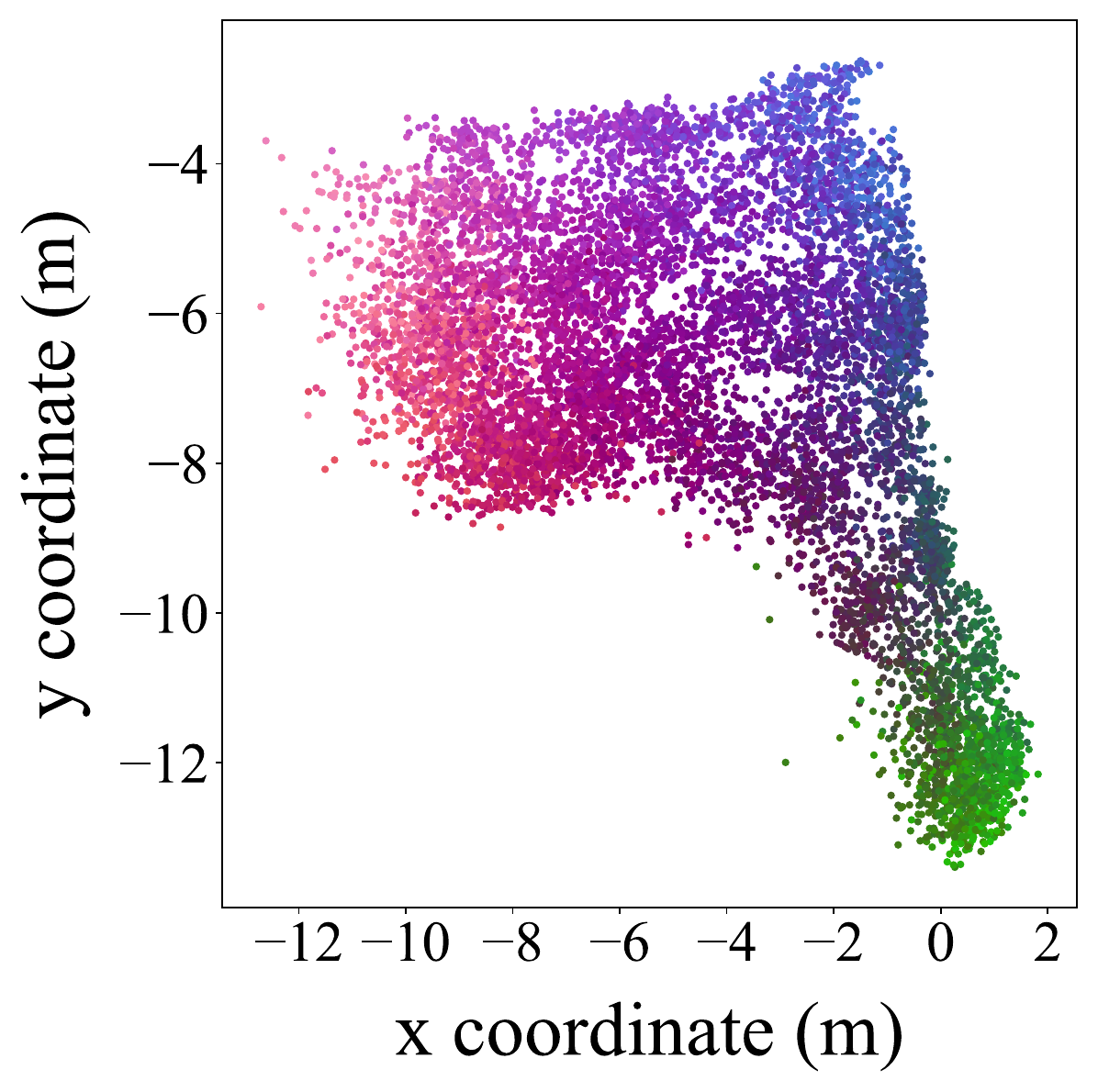}
        \caption{}
        \label{fig:top-right}
    \end{subfigure}
    \hfill
    \begin{subfigure}{0.19\textwidth}
        \includegraphics[width=\linewidth]{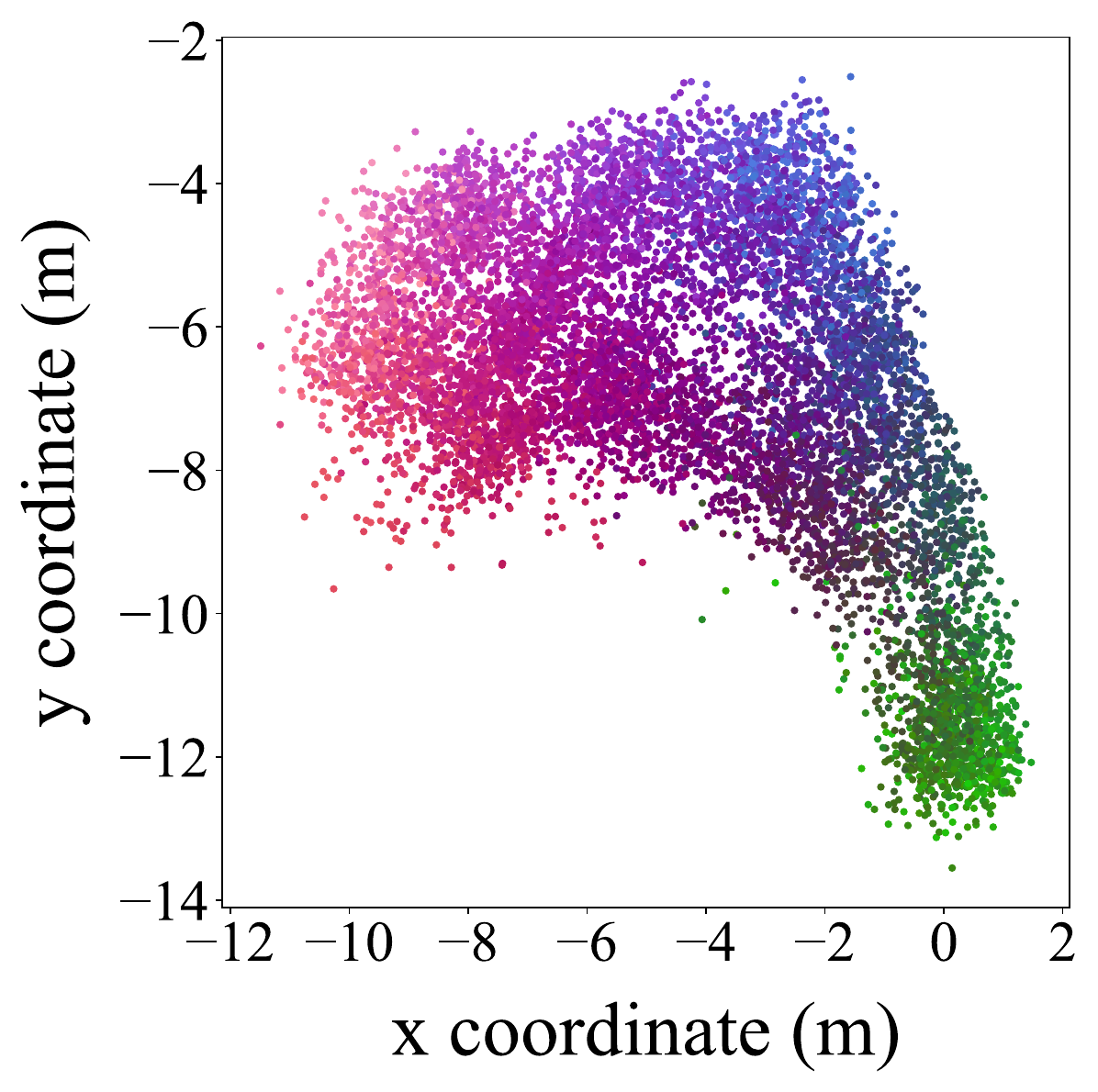}
        \caption{}
        \label{fig:middle}
    \end{subfigure}
    \hfill
    \begin{subfigure}{0.19\textwidth}
        \includegraphics[width=\linewidth]{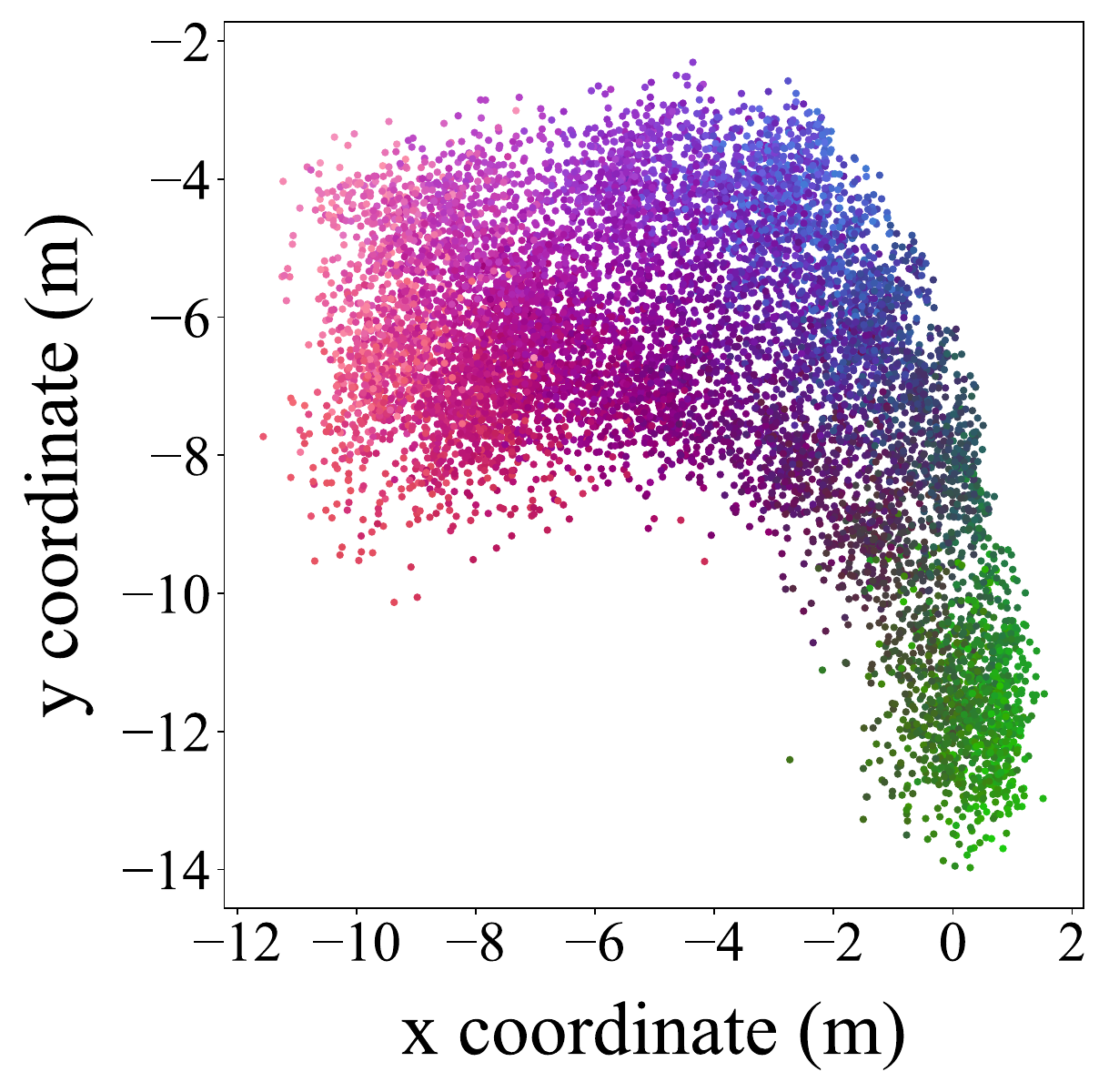}
        \caption{}
        \label{fig:bottom-left}
    \end{subfigure}
    \hfill
    \begin{subfigure}{0.19\textwidth}
        \includegraphics[width=\linewidth]{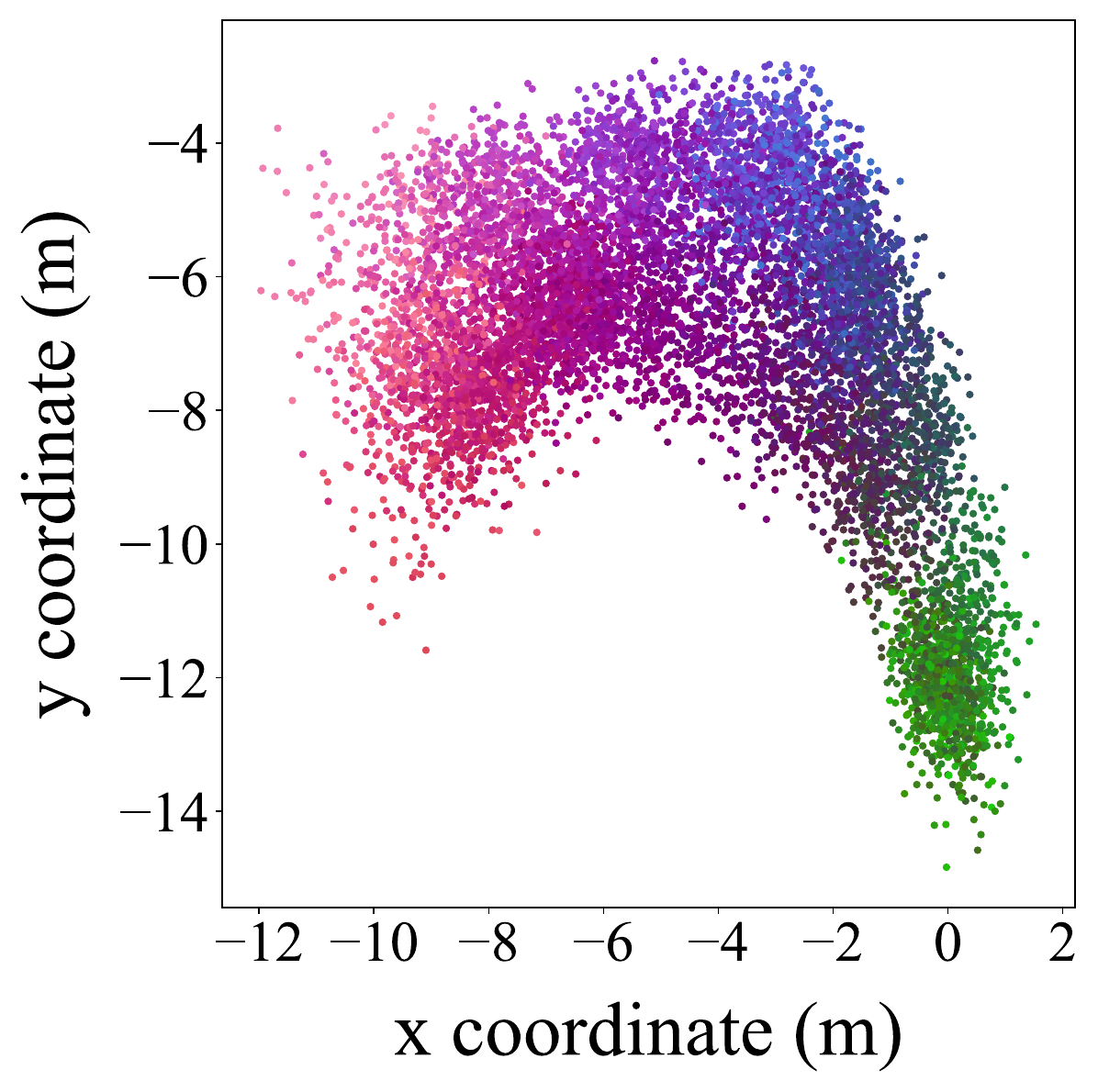} 
        \caption{}
        \label{fig:bottom-right}
    \end{subfigure}
    
    \caption{Inference results based on different channel charts under the scenario of $N_{\text{lab}}$ = 1000 on DICHASUS: (a) Ground truth position; (b)Siamese CC-GAT; (c) Triplet CC-GAT; (d) Isomap CC-GAT; (e) Autoencoder CC-GAT}
    \label{fig:DICHASUS comb}
\end{figure*}

\begin{figure*}[htbp]
    \centering
    \begin{subfigure}{0.19\textwidth}
        \includegraphics[width=\linewidth]{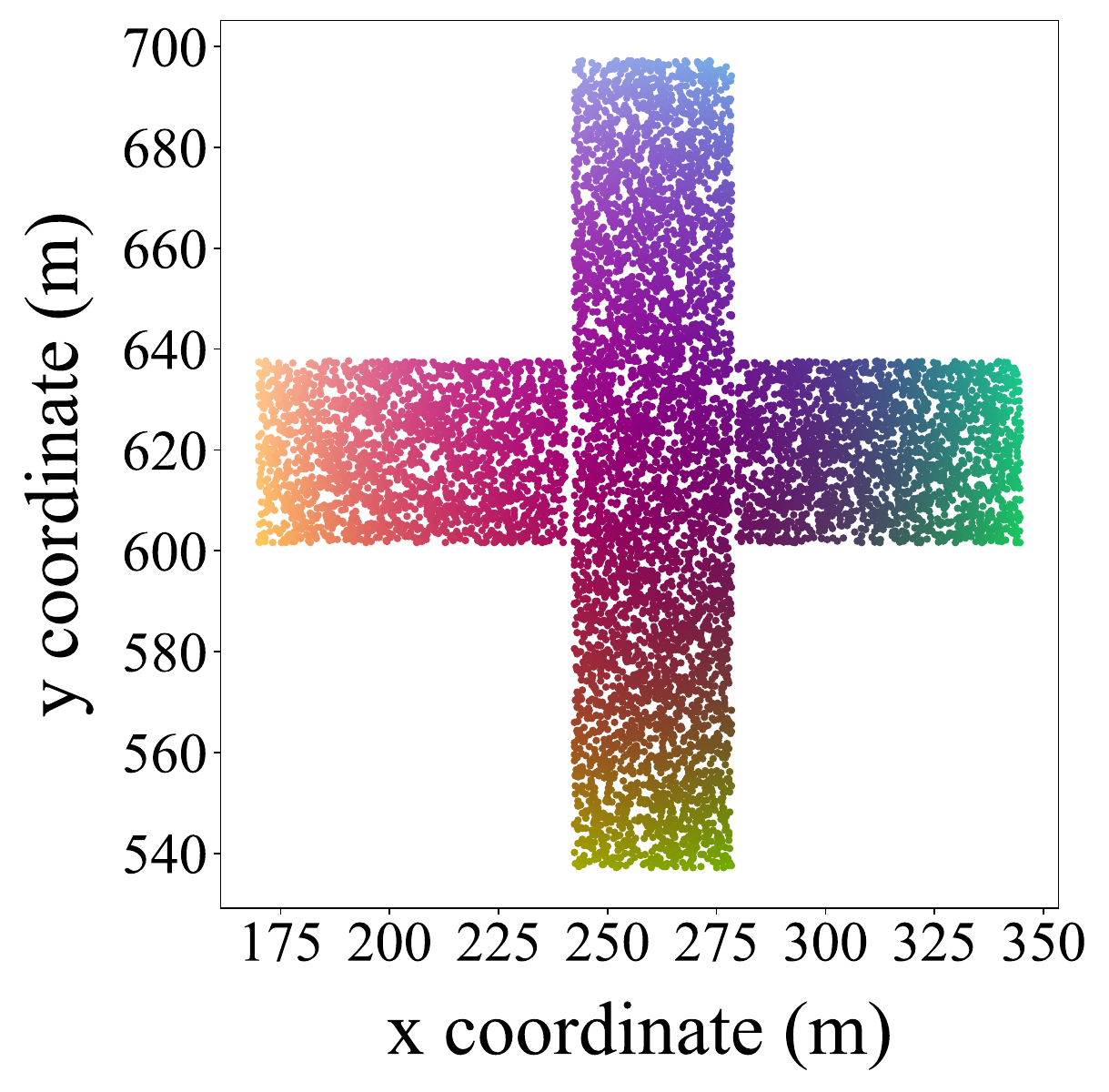}
        \caption{}
        \label{fig:top-left}
    \end{subfigure}
    \hfill
    \begin{subfigure}{0.19\textwidth}
        \includegraphics[width=\linewidth]{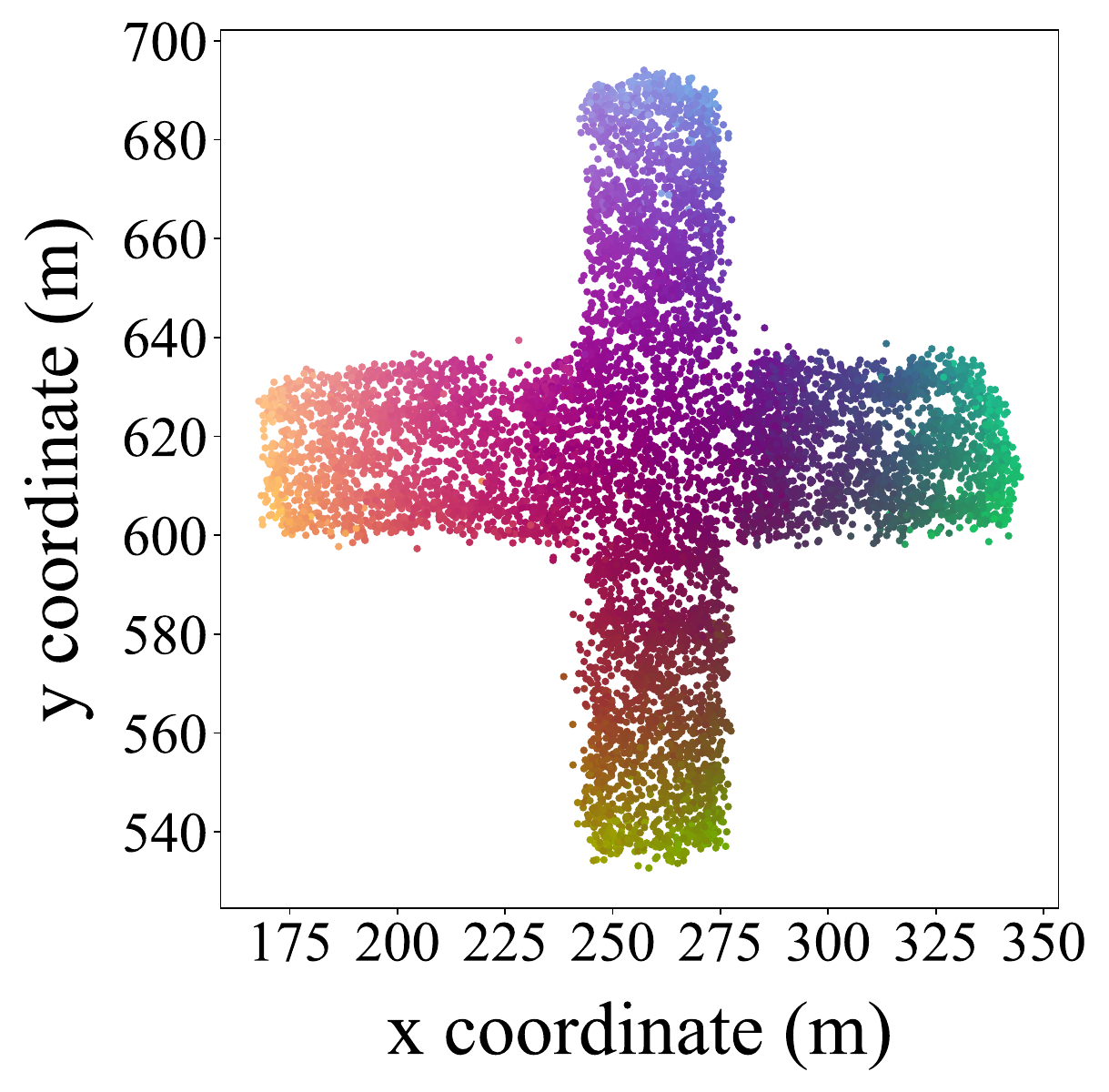}
        \caption{}
        \label{fig:top-right}
    \end{subfigure}
    \hfill
    \begin{subfigure}{0.19\textwidth}
        \includegraphics[width=\linewidth]{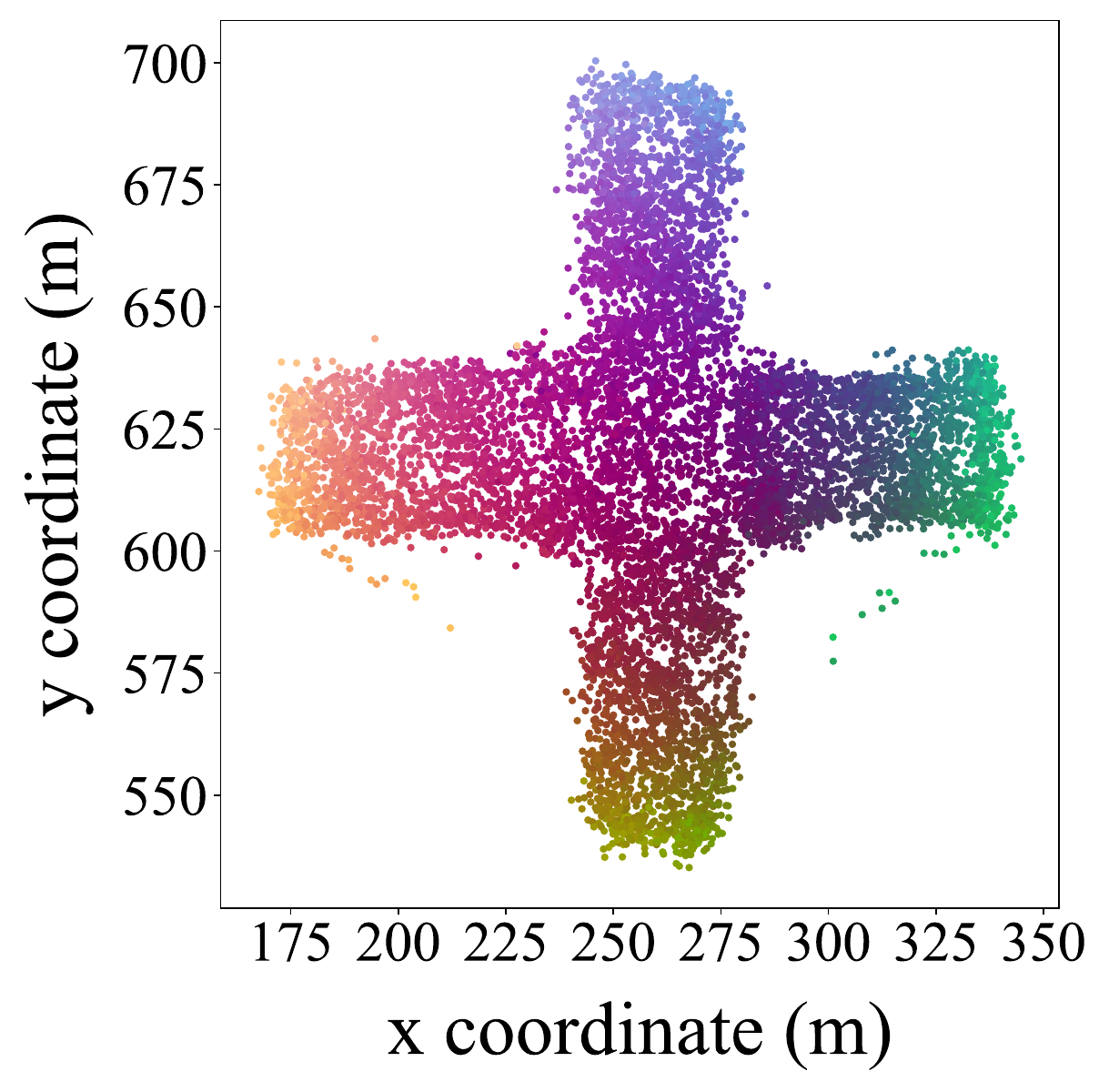}
        \caption{}
        \label{fig:middle}
    \end{subfigure}
    \hfill
    \begin{subfigure}{0.19\textwidth}
        \includegraphics[width=\linewidth]{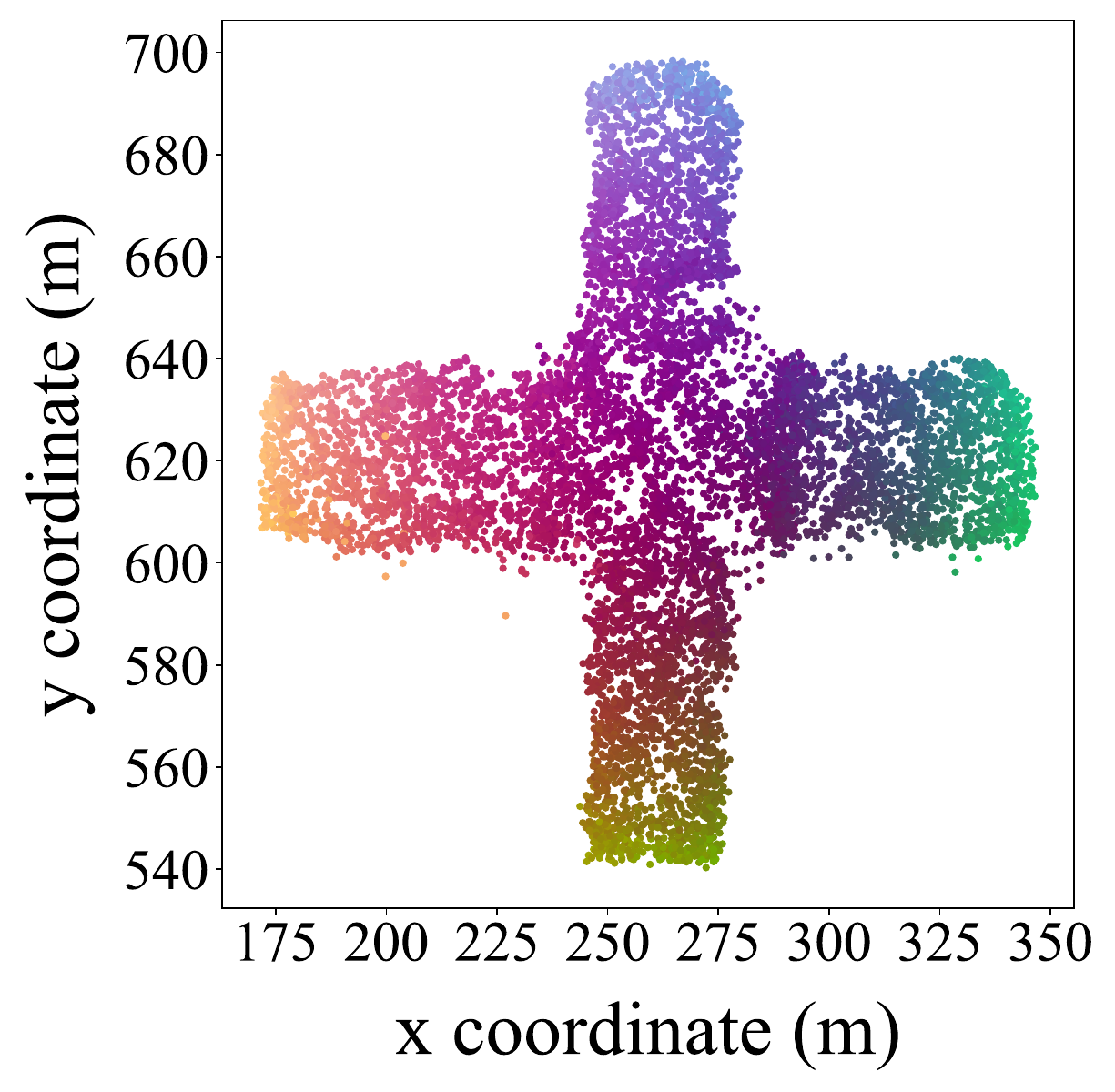}
        \caption{}
        \label{fig:bottom-left}
    \end{subfigure}
    \hfill
    \begin{subfigure}{0.19\textwidth}
        \includegraphics[width=\linewidth]{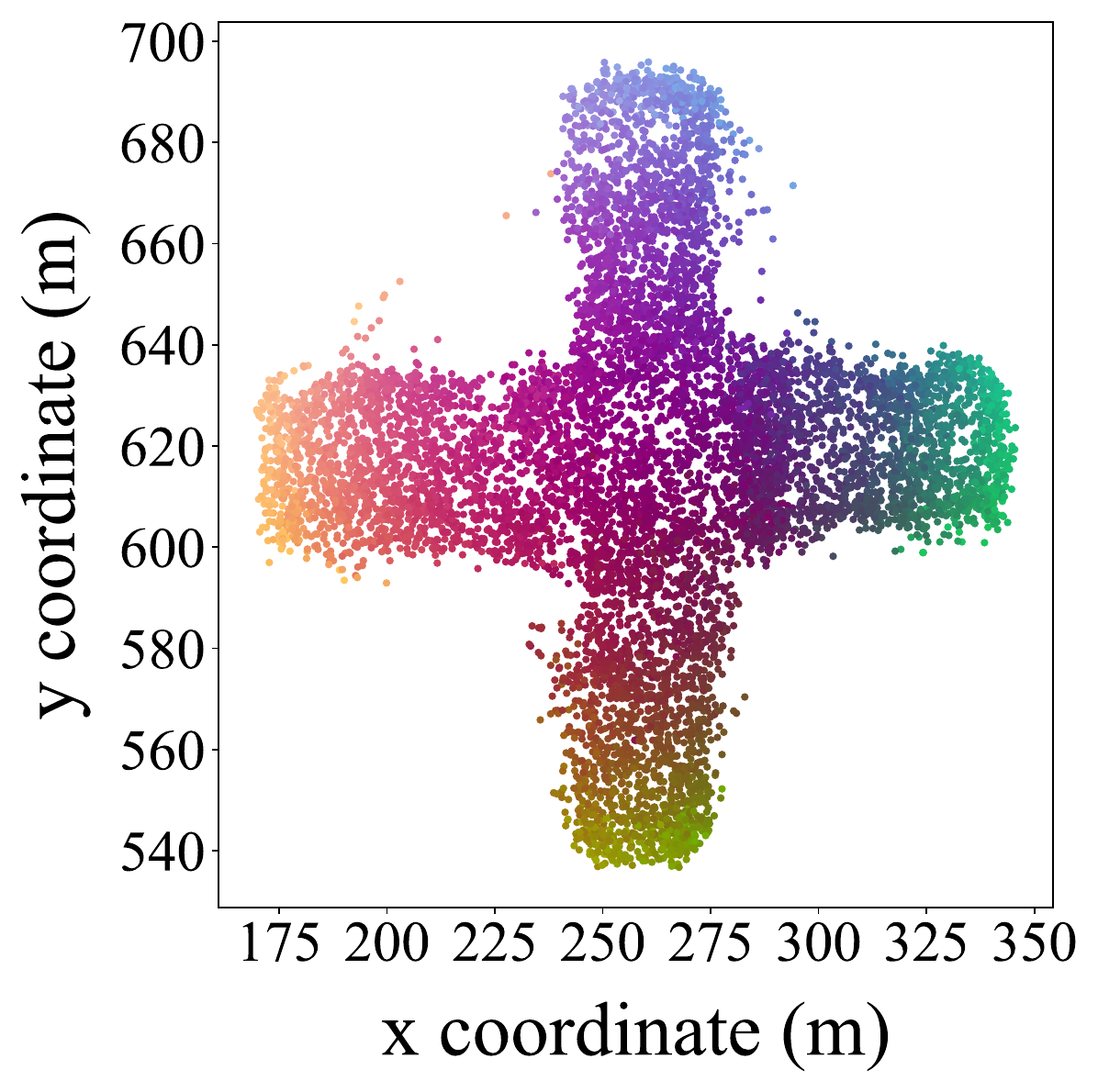} 
        \caption{}
        \label{fig:bottom-right}
    \end{subfigure}
    
    \caption{Inference results based on different channel charts under the scenario of $N_{\text{lab}}$ = 1000 on DeepMIMO: (a) Ground truth position; (b)Siamese CC-GAT; (c) Triplet CC-GAT; (d) Isomap CC-GAT; (e) Autoencoder CC-GAT}
    \label{fig:DeepMIMO comb}
\end{figure*}

\begin{figure}
\centering
\includegraphics[width = 3.5in, height=2.5in]{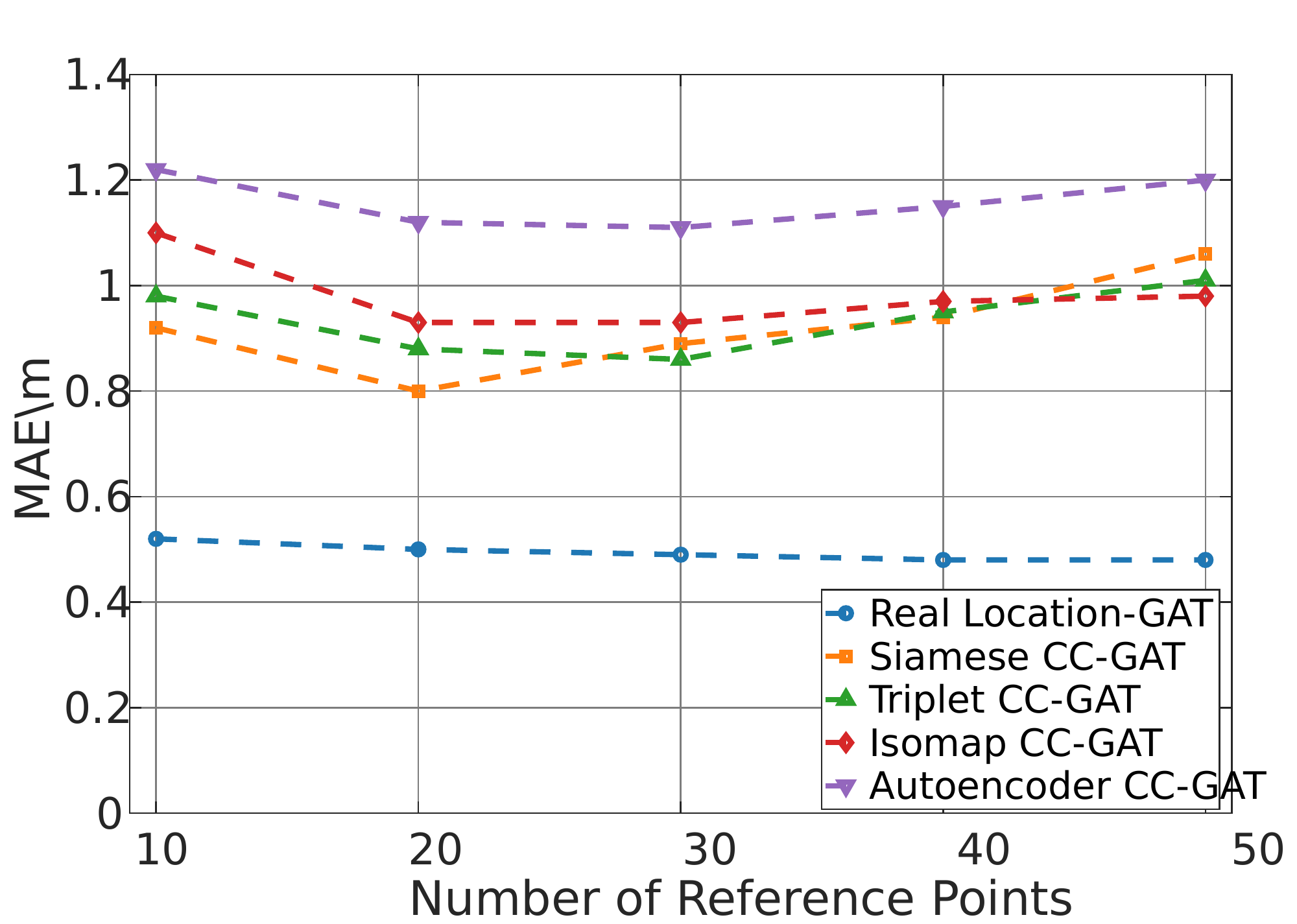}
\caption{MAE performance versus number of RPs on the DICHASUS scene with $N_{\text{lab}}$ = 1000.}
\label{fig:MAE}
\end{figure}
We further visualize the localization results for different CC methods in Fig.~\ref{fig:DICHASUS comb} and Fig.~\ref{fig:DeepMIMO comb}. The point distribution of Siamese CC-GAT shows high consistency with the ground truth positions, whereas Autoencoder CC-GAT noticeably scattered outliers beyond the measurement area. These visual differences corroborate the results of the quantitative analysis, demonstrating that the Siamese method better preserves the spatial topological structure.

The Real Location method establishes a theoretical upper bound, consistently achieving the highest localization accuracy by leveraging perfect position information during RPs selecting. Although this approach is infeasible in practical applications, as it requires knowledge of the unknown query CSI position, it demonstrates the framework's potential with optimal RP selection.
The ADP-based method provides another important benchmark, achieving performance second only to Real Location. Since the ADP dissimilarity metric serves as the training objective for the Siamese and Triplet networks, it represents the optimal similarity measure that these learning-based methods aim to approximate. It should be noted that ADP-GAT, as the theoretical upper bound, is only 7.5\%  better than Siamese CC-GAT, indicating that our learning method is close to the optimal similarity measure. Although ADP-based retrieval offers performance advantages, it incurs significant computational overhead during inference due to the need for high-dimensional CSI similarity calculations. A complete analysis of computational complexity is provided in Sect.~\ref{subsect:complexity}. 
In contrast, learned CC methods achieve efficient inference through pre-computed embeddings, making them more suitable for practical deployment despite the slight performance trade-off.

We further investigate the impact of the number of RPs on the localization performance. As shown in Fig.~\ref{fig:MAE}, all methods exhibit performance improvements with increasing $N_{\text{ref}}$ when $N_{\text{ref}}$ is relatively small. GAT-Real Location maintains superior performance in all configurations, with MAE decreasing from 0.52~m ($N_{\text{ref}}=10$) to 0.48~m ($N_{\text{ref}}=50$), confirming that more RPs provide richer spatial context. However, the performance gain decreases beyond $N_{\text{ref}}=40$, indicating diminishing returns.
On the other hand, all learning-based CC methods exhibit performance degradation when $N_{\text{ref}}$ exceeds 20 or 30. Among all CC approaches, the Siamese CC-GAT achieves optimal performance at $N_{\text{ref}}=20$. Other methods show similar trends, peaking at $N_{\text{ref}}=20$ or 30 before degradation, suggesting that excessive RPs introduce noise that dilutes the informative neighborhood relationships. We recommend $N_{\text{ref}}$ = 20 or 30 for optimal accuracy-efficiency balance, leveraging retrieval-assisted advantages while avoiding noise from excessive RPs.

\begin{figure}
\centering
\includegraphics[width = 3.5in, height=2.5in]{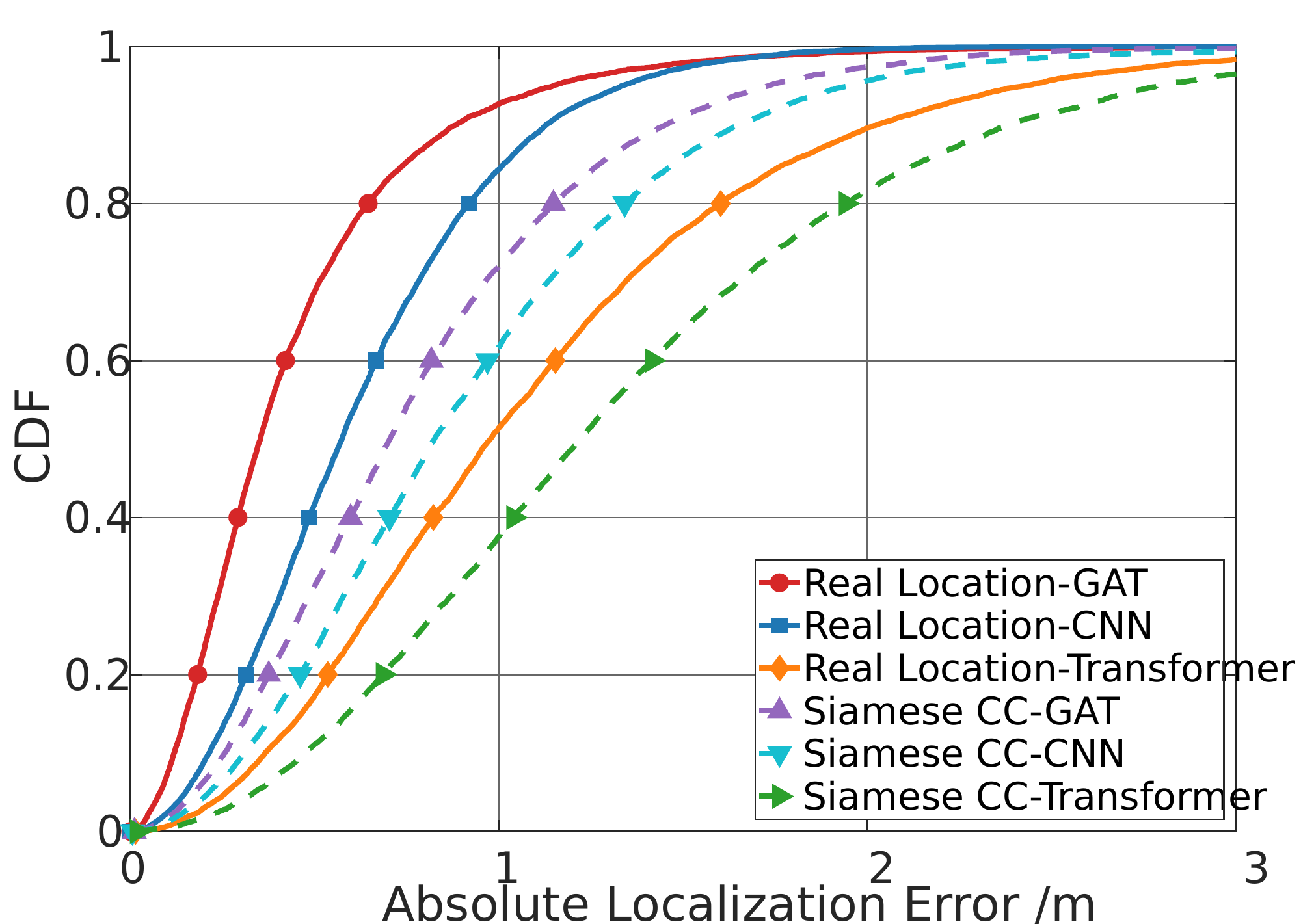}
\caption{The localization error CDF with different localization networks under the scenario of $N_{\text{lab}}$ = 1000 on the DICHASUS scene.}
\label{fig:CDF_LOC}
\end{figure}

\subsection{Performance Comparison of Localization Networks}
\label{subsec:network_comparison}

Fig.~\ref{fig:CDF_LOC} presents the CDF comparison of different localization networks under the scenario of $N_{\text{lab}}=1000$ on the DICHASUS scene. The results demonstrate the consistent superiority of GAT-based localization in both theoretical and practical settings.
When employing Real Location as the retrieval strategy, which represents the theoretical upper bound of our framework, GAT achieves the best performance with 50\% and 90\% localization errors of 0.35~m and 0.88~m, respectively. This significantly outperforms both CNN-based and Transformer-based approaches under the same ideal conditions. More importantly, in the practical application scenario using Siamese CC for retrieval, GAT maintains its advantage with 50\% and 90\% errors of 0.70~m and 1.45~m, representing approximately 13.9\% improvement over CNN and 39.3\% over Transformer.

The superior performance of GAT stems from its inherent architectural advantages in processing graph-structured data, which fundamentally differs from the suboptimal input representations used by CNN and Transformer. Unlike CNN and Transformer, which simply stack all the features along the RP dimension, treating them as an unordered collection, GAT explicitly models pairwise relationships through its attention mechanism, enabling adaptive, correlation-aware aggregation that dynamically weights each RP based on its relevance to the query sample. This attention mechanism allows GAT to emphasize informative references while suppressing noisy or less relevant ones, effectively leveraging the graph structure inherent in retrieval-based localization. In contrast, CNN's fixed geometric inductive bias and Transformer's sequence-oriented approach are less suited for handling unordered reference sets where relational information is crucial. The consistent outperformance of GAT across both theoretical and practical scenarios validates that explicit graph structure modeling offers significant benefits over the simple input stacking strategies employed by other network architectures.

\subsection{Complexity Comparison}
\label{subsect:complexity}

This section provides a computational complexity analysis of the proposed framework, focusing on the efficiency advantages of channel charting-based retrieval.
The ADP-based retrieval method requires computing cosine similarities across all dimensions of the CSI. For a database of size $N_{\text{lab}}$ , the overall computational complexity is $\mathcal{O}$($N_{\text{AP}} \cdot N_r \cdot N_{\text{BS}} \cdot N_{\text{lab}}$). Taking the DICHASUS scenario with $N_{\text{lab}}$ = 1000 as an example, this results in 4×8×1024×1000 = 32,768,000 dimensional computations, which becomes prohibitively expensive for real-time applications. Furthermore, this complexity increases linearly with $N_{\text{lab}}$, imposing additional overhead on continuous learning systems that require periodic expansion of the fingerprint database.

In contrast,  the proposed CC-based approach projects high-dimensional CSI data into a low-dimensional embedding space, reducing the similarity computation complexity to $\mathcal{O}$($N_{\text{CC}}\cdot N_{\text{lab}}$), where $N_{\text{CC}}$ denotes the dimensionality of the learned channel chart. As the database grows, the linear complexity $\mathcal{O}$($N_{\text{CC}}\cdot N_{\text{lab}}$) ensures feasible operation in real-time systems, whereas the high complexity of ADP-based search becomes a critical bottleneck. This represents a substantial improvement in computational efficiency.

\begin{table}[htbp]
\centering
\caption{Computational efficiency comparison with $N_{\text{lab}}$ = 1000 on DICHASUS.}
\label{tab:efficiency_comparison}
\begin{tabular}{@{}l@{\hspace{0.4em}}c@{\hspace{0.4em}}c@{}}
\toprule
\textbf{Method} & \textbf{Construct Time (ms)} & \textbf{Inference Time (ms)} \\
\midrule
CC-GAT & $1.1 \pm 0.2$ & $1.3 \pm 0.2$  \\
ADP-GAT & $212.5 \pm 10.2 $ & $1.2 \pm 0.2$  \\
CC-CNN & $1.1 \pm 0.2$ & $0.4 \pm 0.1$ \\
CC-Transformer & $1.1 \pm 0.2$ & $0.6 \pm 0.1$ \\
WKNN & - & $2.1 \pm 0.3$ \\
\bottomrule
\end{tabular}
\end{table}
Table~\ref{tab:efficiency_comparison} presents empirical timing results, demonstrating that CC-based retrieval provides approximately 100 times faster reference point selection compared to ADP-based methods. This significant efficiency advantage, combined with competitive localization accuracy, makes the CC-based approach more suitable for practical deployment scenarios.

\section{Conclusion} \label{sect:conclusion}

This paper proposed a unified retrieval-assisted fingerprinting localization framework that integrates similarity-based reference point retrieval with learning-based localization inference. By explicitly incorporating retrieved reference fingerprints during inference, the proposed approach overcomes the scalability limitations of traditional similarity-based methods and the generalization issues of purely learning-based localization models. Specifically, channel charting is employed to enable efficient retrieval of locally correlated reference points, while a graph attention network is used to model inter-sample correlations and perform adaptive feature aggregation for accurate position estimation. Extensive experiments on real-world indoor measurements and ray-tracing-based outdoor datasets demonstrate that the proposed framework consistently outperforms state-of-the-art localization methods. Future work will consider extending the proposed framework to more dynamic environments, such as online reference database updates and environment-aware adaptation. In addition, exploring conceptual connections between retrieval-assisted localization and retrieval-augmented inference paradigms may also be an interesting direction for future research.
Furthermore, exploring joint end-to-end optimization may further enhance performance and represent an interesting direction for research.

\bibliographystyle{IEEEtran}
\bibliography{IEEEabrv,bb_rf}

\begin{thebibliography}{10}
\providecommand{\url}[1]{#1}
\csname url@samestyle\endcsname
\providecommand{\newblock}{\relax}
\providecommand{\bibinfo}[2]{#2}
\providecommand{\BIBentrySTDinterwordspacing}{\spaceskip=0pt\relax}
\providecommand{\BIBentryALTinterwordstretchfactor}{4}
\providecommand{\BIBentryALTinterwordspacing}{\spaceskip=\fontdimen2\font plus
\BIBentryALTinterwordstretchfactor\fontdimen3\font minus \fontdimen4\font\relax}
\providecommand{\BIBforeignlanguage}[2]{{%
\expandafter\ifx\csname l@#1\endcsname\relax
\typeout{** WARNING: IEEEtran.bst: No hyphenation pattern has been}%
\typeout{** loaded for the language `#1'. Using the pattern for}%
\typeout{** the default language instead.}%
\else
\language=\csname l@#1\endcsname
\fi
#2}}
\providecommand{\BIBdecl}{\relax}
\BIBdecl

\bibitem{pan2025ai}
G.~Pan, Y.~Gao, Y.~Gao, W.~Yu, Z.~Zhong, X.~Yang, X.~Guo, and S.~Xu, ``{AI}-driven wireless positioning: Fundamentals, standards, state-of-the-art, and challenges,'' \emph{IEEE Commun. Surveys Tuts.}, vol.~28, pp. 4394--4428, 2026.

\bibitem{Zafari2019indoor}
F.~Zafari, A.~Gkelias, and K.~K. Leung, ``A survey of indoor localization systems and technologies,'' \emph{IEEE Commun. Surveys Tuts.}, vol.~21, no.~3, pp. 2568--2599, 2019.

\bibitem{Henk2025}
H.~Wymeersch, H.~Chen, H.~Guo, M.~F. Keskin, B.~M. Khorsandi, M.~H. Moghaddam, A.~Ramirez, K.~Schindhelm, A.~Stavridis, T.~Svensson, and V.~Yajnanarayana, ``6{G} positioning and sensing through the lens of sustainability, inclusiveness, and trustworthiness,'' \emph{IEEE Wireless Commun.}, vol.~32, no.~1, pp. 68--75, 2025.

\bibitem{yang2024positioning}
Y.~Yang, M.~Chen, Y.~Blankenship, J.~Lee, Z.~Ghassemlooy, J.~Cheng, and S.~Mao, ``Positioning using wireless networks: Applications, recent progress and future challenges,'' \emph{IEEE J. Sel. Areas Commun.}, vol.~42, no.~9, pp. 2149--2178, 2024.

\bibitem{8515231}
R.~Mendrzik, H.~Wymeersch, G.~Bauch, and Z.~Abu-Shaban, ``Harnessing {NLOS} components for position and orientation estimation in {5G} millimeter wave {MIMO},'' \emph{IEEE Trans. Wireless Commun.}, vol.~18, no.~1, pp. 93--107, 2019.

\bibitem{Yang2024}
S.~Yang, D.~Zhang, R.~Song, P.~Yin, and Y.~Chen, ``Multiple {WiFi} access points co-localization through joint {AoA} estimation,'' \emph{IEEE Trans. Mobile Comput.}, vol.~23, no.~2, pp. 1488--1502, 2024.

\bibitem{Italiano5GTutorial}
L.~Italiano, B.~Camajori~Tedeschini, M.~Brambilla, H.~Huang, M.~Nicoli, and H.~Wymeersch, ``A tutorial on 5{G} positioning,'' \emph{IEEE Commun. Surveys Tuts.}, vol.~27, no.~3, pp. 1488--1535, 2025.

\bibitem{zhang2025leveraging}
J.~Zhang, J.~Xue, Y.~Li, and S.~L. Cotton, ``Leveraging online learning for domain-adaptation in {Wi-Fi}-based device-free localization,'' \emph{IEEE Trans. Mobile Comput.}, 2025.

\bibitem{chen2022tutorial}
H.~Chen, H.~Sarieddeen, T.~Ballal, H.~Wymeersch, M.-S. Alouini, and T.~Y. Al-Naffouri, ``A tutorial on terahertz-band localization for {6G} communication systems,'' \emph{IEEE Commun. Surveys Tuts.}, vol.~24, no.~3, pp. 1780--1815, 2022.

\bibitem{Sadowski_2020_KNN}
S.~Sadowski, P.~Spachos, and K.~N. Plataniotis, ``Memoryless techniques and wireless technologies for indoor localization with the internet of things,'' \emph{IEEE Internet Things J.}, vol.~7, no.~11, pp. 10\,996--11\,005, 2020.

\bibitem{KNN2022Ali}
K.~Ali and A.~X. Liu, ``Fine-grained vibration based sensing using a smartphone,'' \emph{IEEE Trans. Mobile Comput.}, vol.~21, no.~11, pp. 3971--3985, 2022.

\bibitem{Hu_2024_KNN}
Z.~Hu, X.~Chen, Z.~Zhou, and S.~Mumtaz, ``Localization with cellular signal {RSRP} fingerprint of multiband and multicell,'' \emph{IEEE J. Sel. Areas Commun.}, vol.~42, no.~9, pp. 2380--2394, 2024.

\bibitem{luckner2023selection}
M.~Luckner, S.~Sowik, and P.~Brida, ``Selection of signal sources influence at indoor positioning system,'' \emph{IEEE Trans. Wireless Commun.}, vol.~23, no.~1, pp. 45--57, 2023.

\bibitem{Lin2025_MLP}
L.~Li, Z.~Xing, X.~Guo, and H.~Zheng, ``D²-mloc: Dual-domain mlp-mixer framework for {CSI}-fingerprinting indoor localization,'' \emph{IEEE Trans. Cogn. Commun. Netw.}, vol.~11, no.~5, pp. 3276--3291, 2025.

\bibitem{GAO_2024CNN}
R.~Gao, S.~Zhu, L.~Li, X.~Wang, Y.~Jiang, N.~Tan, H.~Chai, P.~Qi, J.~Liu, and D.~Tao, ``Real-world large-scale cellular localization for pickup position recommendation at black-hole,'' \emph{IEEE Trans. Mobile Comput.}, vol.~23, no.~12, pp. 15\,114--15\,131, 2024.

\bibitem{zhang2023csi}
B.~Zhang, H.~Sifaou, and G.~Y. Li, ``{CSI}-fingerprinting indoor localization via attention-augmented residual convolutional neural network,'' \emph{IEEE Trans. Wireless Commun.}, vol.~22, no.~8, pp. 5583--5597, 2023.

\bibitem{pan2025large}
G.~Pan, K.~Huang, H.~Chen, S.~Zhang, C.~H{\"a}ger, and H.~Wymeersch, ``Large wireless localization model {(LWLM)}: A foundation model for positioning in {6G} networks,'' \emph{arXiv preprint arXiv:2505.10134}, 2025.

\bibitem{Zhang2023_GNN1}
M.~Zhang, Z.~Fan, R.~Shibasaki, and X.~Song, ``Domain adversarial graph convolutional network based on rssi and crowdsensing for indoor localization,'' \emph{IEEE Internet Things J.}, vol.~10, no.~15, pp. 13\,662--13\,672, 2023.

\bibitem{Ye_2025_GNN2}
Z.~Ye, Q.~Xiao, J.~Liu, Y.~He, G.~Yu, and J.~Han, ``Practical {WiFi} indoor localization: Unleashing the potential of gnns for accuracy and robustness,'' \emph{IEEE Trans. Mobile Comput.}, pp. 1--17, 2025.

\bibitem{GCNLoc}
Y.~Sun, Q.~Xie, G.~Pan, S.~Zhang, and S.~Xu, ``A novel {GCN} based indoor localization system with multiple access points,'' in \emph{Proc. IEEE IWCMC}, 2021, pp. 9--14.

\bibitem{Zhu2023}
X.~Zhu, T.~Qiu, W.~Qu, X.~Zhou, M.~Atiquzzaman, and D.~O. Wu, ``{BLS-L}ocation: A wireless fingerprint localization algorithm based on broad learning,'' \emph{IEEE Trans. Mobile Comput.}, vol.~22, no.~1, pp. 115--128, 2023.

\bibitem{stahlke2023indoor}
M.~Stahlke, G.~Yammine, T.~Feigl, B.~M. Eskofier, and C.~Mutschler, ``Indoor localization with robust global channel charting: A time-distance-based approach,'' \emph{IEEE Trans. Mach. Learn. Commun. Netw.}, vol.~1, pp. 3--17, 2023.

\bibitem{stephan2024angle}
P.~Stephan, F.~Euchner, and S.~Ten~Brink, ``Angle-delay profile-based and timestamp-aided dissimilarity metrics for channel charting,'' \emph{IEEE Trans. Commun.}, vol.~72, no.~9, pp. 5611--5625, 2024.

\bibitem{ferrand2021triplet}
P.~Ferrand, A.~Decurninge, L.~G. Ordonez, and M.~Guillaud, ``Triplet-based wireless channel charting: Architecture and experiments,'' \emph{IEEE J. Sel. Areas Commun.}, vol.~39, no.~8, pp. 2361--2373, 2021.

\bibitem{taner2025channel}
S.~Taner, V.~Palhares, and C.~Studer, ``Channel charting in real-world coordinates with distributed {MIMO},'' \emph{IEEE Trans. Wireless Commun.}, 2025.

\bibitem{mateos2025positioning}
J.~M. Mateos-Ramos, F.~Zumegen, H.~Wymeersch, C.~H{\"a}ger, and C.~Studer, ``Positioning via digital-twin-aided channel charting with large-scale {CSI} features,'' \emph{arXiv preprint arXiv:2511.09227}, 2025.

\bibitem{zhang2025unilocpro}
Y.~Zhang, G.~Pan, M.~F. Keskin, O.~Kaltiokallio, M.~Valkama, and H.~Wymeersch, ``{UNILocPro}: Unified localization integrating model-based geometry and channel charting,'' \emph{arXiv preprint arXiv:2510.27394}, 2025.

\bibitem{Yuan2025}
X.~Yuan, M.~Zhang, Y.~Zheng, B.~Teng, and W.~Jiang, ``Scalable near-field localization based on partitioned large-scale antenna array,'' \emph{IEEE Trans. Wireless Commun.}, vol.~24, no.~3, pp. 2203--2217, 2025.

\bibitem{zhou2021integrated}
M.~Zhou, Y.~Li, M.~J. Tahir, X.~Geng, Y.~Wang, and W.~He, ``Integrated statistical test of signal distributions and access point contributions for {Wi-Fi} indoor localization,'' \emph{IEEE Trans. Veh. Technol.}, vol.~70, no.~5, pp. 5057--5070, 2021.

\bibitem{hu2018experimental}
J.~Hu, D.~Liu, Z.~Yan, and H.~Liu, ``Experimental analysis on weight k-nearest neighbor indoor fingerprint positioning,'' \emph{IEEE Internet Things J.}, vol.~6, no.~1, pp. 891--897, 2018.

\bibitem{velickovic2017graph}
P.~Velickovic, G.~Cucurull, A.~Casanova, A.~Romero, P.~Lio, Y.~Bengio \emph{et~al.}, ``Graph attention networks,'' in \emph{Proc. ICLR'2017}, vol. 1050, no.~20, 2017, pp. 10--48\,550.

\bibitem{yapar2023real}
{\c{C}}.~Yapar, R.~Levie, G.~Kutyniok, and G.~Caire, ``Real-time outdoor localization using radio maps: A deep learning approach,'' \emph{IEEE Trans. Wireless Commun.}, vol.~22, no.~12, pp. 9703--9717, 2023.

\bibitem{he2016deep}
K.~He, X.~Zhang, S.~Ren, and J.~Sun, ``Deep residual learning for image recognition,'' in \emph{Proc. IEEE CVPR'2016}, 2016, pp. 770--778.

\bibitem{euchner2021distributed}
F.~Euchner, M.~Gauger, S.~D{\"o}rner, and S.~ten Brink, ``A distributed massive {MIMO} channel sounder for “big {CSI} data”-driven machine learning,'' in \emph{Proc. WSA'2021}.\hskip 1em plus 0.5em minus 0.4em\relax VDE, 2021, pp. 1--6.

\bibitem{alkhateeb2019deepmimo}
A.~Alkhateeb, ``{DeepMIMO}: A generic deep learning dataset for millimeter wave and massive {MIMO} applications,'' \emph{arXiv preprint arXiv:1902.06435}, 2019.

\bibitem{Clevert2015}
D.-A. Clevert, T.~Unterthiner, and S.~Hochreiter, ``Fast and accurate deep network learning by exponential linear units ({ELUs}),'' in \emph{Proc. ICLR'2016}, 2016.

\bibitem{dataset-dichasus-cf0x}
\BIBentryALTinterwordspacing
F.~Euchner and M.~Gauger, ``{CSI Dataset dichasus-cf0x: Distributed Antenna Setup in Industrial Environment, Day 1},'' 2022. [Online]. Available: \url{https://doi.org/doi:10.18419/darus-2854}
\BIBentrySTDinterwordspacing

\bibitem{vaswani2017attention}
A.~Vaswani, N.~Shazeer, N.~Parmar, J.~Uszkoreit, L.~Jones, A.~N. Gomez, {\L}.~Kaiser, and I.~Polosukhin, ``Attention is all you need,'' in \emph{Proc. NIPS’2017}, vol.~30, 2017.

\bibitem{gong2023deep}
X.~Gong, A.~Lu, X.~Liu, X.~Fu, X.~Gao, and X.-G. Xia, ``Deep learning based fingerprint positioning for multi-cell massive {MIMO-OFDM} systems,'' \emph{IEEE Trans. Veh. Technol.}, vol.~73, no.~3, pp. 3832--3849, 2023.

\bibitem{tenenbaum2000global}
J.~B. Tenenbaum, V.~d. Silva, and J.~C. Langford, ``A global geometric framework for nonlinear dimensionality reduction,'' \emph{science}, vol. 290, no. 5500, pp. 2319--2323, 2000.

\end{thebibliography}

\end{document}